\documentclass[12pt]{iopart}
\pdfoutput=1
\usepackage{iopams}
\usepackage{graphicx}
\usepackage{cite}
\renewcommand{\vec}[1]{\mbox{\protect\boldmath$#1$}}

\begin{document}
\title[Regular and chaotic orbits near a massive magnetic dipole]
{Regular and chaotic orbits near a massive magnetic dipole}

\author{Ji\v{r}\'{\i} Kov\'{a}\v{r}$^1\!\!$, Ond\v{r}ej Kop\'{a}\v{c}ek$^2\!\!$, Vladim\'{\i}r Karas$^2$ and Yasufumi Kojima$^3$}
\address{
$^1$~Institute of Physics, Faculty of Philosophy and Science, Silesian University in Opava, Bezru\v{c}ovo n\'{a}m.~13, CZ-746\,01~Opava, Czech~Republic\\
$^2$~Astronomical Institute, Academy of Sciences, Bo\v{c}n\'{i} II, CZ-141\,31~Prague, Czech~Republic\\
$^3$~Department of Physics, Hiroshima University, Higashi-Hiroshima 739-8526, Japan}
\ead{jiri.kovar@fpf.slu.cz, kojima@theo.phys.sci.hiroshima-u.ac.jp}

\begin{abstract}
Within the framework of Bonnor's exact solution describing a massive magnetic dipole, we study the motion of neutral and electrically charged test particles. In dependence on the Bonnor spacetime parameters, we determine regions enabling the existence of stable circular orbits confined to the equatorial plane and of those levitating above the equatorial plane. Constructing Poincar\'{e} surfaces of section and recurrence plots, we also investigate the dynamics of particles moving along general off-equatorial trajectories bound in effective potential wells forming around the stable circular orbits. We demonstrate that the motion in the Bonnor spacetime is not integrable. This extends previous investigations of generalized St\"ormer's problem into the realm of exact solutions of Einstein-Maxwell equations, where the gravitational and electromagnetic effects play a comparable role on the particle motion.  
\end{abstract}


\maketitle

\section{\label{Sec:Intro}Introduction}

Since the seminal work by St\"ormer \cite{Stormer55}, many authors have investigated properties of motion of electrically charged 
particles through a dipole-type magnetic field as a model for magnetospheres surrounding planets. The original motivation
of the early works was to understand the trajectories and radiation of plasma particles which arrive with the solar wind and become 
captured by the Earth magnetic field. Indeed, radiation belts are now known to be formed by ions and electrons whose motion 
is governed by magnetic forces. Beside these classical studies, the St\"ormer's work was used in the context of General Relativity to study the stable orbit around Kerr black hole by Ruffini and Wheeler. Their work is reproduced among others in figure 18 of the book \cite{Rees74} and has been quoted in \cite{Landau75} in paragraph 104. All the definitions of the stable and unstable orbits were there basically obtained by using the St\"ormer's approach in General Relativity (see also \cite{mtw}).

Later on, different generalizations of the original St\"ormer's discussion have been developed in order to describe various aspects of the 
problem and, to capture more complicated situations \cite{Howard99,Howard01}. For example, in a case of dust particles, only a weak electric charge is established on the surface of the particles, i.e., small charge to mass ratio develops by photoionization 
or selective charge accretion \cite{Vladimirov05}. Therefore, one needs to take also gravitational force into account. At the same 
time the influence of rotation induced electric field becomes non-negligible and has to be combined with the magnetic component.
This is the essence of the so-called generalized St\"ormer problem \cite{Dullin02,Howard99}. Total gravitational and electromagnetic forces then determine regions of stable motion, in particular, whether off-equatorial potential wells (halo lobes) develop where particles can be captured in permanent circulation around the central body. The problem is relevant in the context of accreting compact objects, where the motion within the halo lobes (halo motion) can describe the overall global motion of plasma particles through corona of an accretion disc.

Relativistic effects change the regions of stability near a compact star or a black hole. In order to reveal the existence of the halo motion under strong gravity regime, we adopt the formalism of General Relativity.
To this end, a slowly rotating compact star endowed with an aligned magnetic dipolar field, or a magnetized black hole embedded in an external magnetic field of a current loop outside the horizon were previously studied \cite{Kovar08,Kovar10,Takahashi09}. 
Furthermore, an asymptotically uniform Wald magnetic field parallel to the rotation axis of Kerr 
black hole was also considered \cite{Kovar10,kopacek10}. Regions of stable circular halo motion were found for suitable combinations of model parameters, such as the generalized energy and angular momentum of the particles, their specific electric charge, the magnetic dipole moment of the central body (defining the intensity of the large scale magnetic field), and angular velocity of rotation. Note that an approximative analysis of the halo motion in strong gravitational fields can be also formulated by employing the pseudo-Newtonian approach (see, e.g., paper \cite{Semerak99} for an overview and references). Although this method does not allow a mathematically rigorous discussion of the problem, it can provide useful clues about the effects and, indeed, it was found \cite{Kovar08} that the existence and position of the stable halo circular orbits of charged particles are reproduced very accurately by the Paczy\'nski \& Wiita model.

Whilst the above-mentioned examples include gravitational effects by adopting the metric of a black hole, the electromagnetic 
influence has been so far considered in terms of only a weak (test) field approximation. The Kerr-Newman black hole offers a way to study the 
problem within the framework of an exact solution of combined Einstein-Maxwell equations. However, this particular example turns 
out to be quite uninteresting, because the halo lobes were not found above the outer horizon \cite{Kovar08}. 
The lack of stable circular halo orbits outside the horizon of Kerr-Newman black hole is not very surprising (it can be ascribed 
to the special character of this solution), nonetheless, we want to explore the occurrence of these orbits and study the related halo motion in an exact solution that can be used to describe a massive magnetic dipole as the closest analogy to the Newtonian generalized St\"{o}rmer problem. To do this, we employ the Bonnor spacetime as the suitable example of such a background. 

Test particles motion in strong gravitational and electromagnetic fields has been ranked among basic problems in relativistic astrophysics, widely studied since the conception of General Relativity, still having high theoretical, practical and educational significance (see, e.g., the recent nice serie of papers \cite{Pugliese11a,Pugliese11b,Pugliese11c} or the paper \cite{Bakala12} and the reference therein).

%%%%%%%%%%%%%%%%%%%%%%%%%%%%%%%%%%%%%%%%%%%%%%%%%%%%%%%%%%%%%%%%%%%%%%%%%%%%%%%%%%%%%%%%%%%%%%%%%%
\section{Bonnor's solution and equations of motion}
Bonnor derived a special solution \cite{Bonnor66} of combined Einstein-Maxwell equations describing a static massive source carrying a magnetic dipol. In  $(t,r,\theta,\phi)$ coordinates and $c=G=1$ units the solution reads  
\begin{eqnarray}
\label{metric}
{\rm d}s^2&=&-\left(\frac{P}{Y}\right)^2 {\rm d}t^2 +\frac{P^2Y^2}{Q^3Z}({\rm d}r^2+Z{\rm d}\theta^2)
+\frac{Y^2Z\sin^2{\theta}}{P^2}{\rm d}\phi^2,\\
\label{EMpot}
A_{\alpha}&=&(0,0,0,\frac{\mu r\sin^2{\theta}}{P}),
\end{eqnarray}
where we adopted the notation
\begin{eqnarray}
P&=&r^2-2ar-b^2\cos^2{\theta}, \\
Q&=&(r-a)^2 -(a^2+b^2)\cos^2{\theta},\\
Y&=&r^2-b^2\cos^2{\theta}, \\
Z&=&r^2-2ar-b^2.
\end{eqnarray}
The solution is asymptotically flat (for $a=0$ exactly flat) and of Weyl tensor type [1111]. It is characterized by two independent parameters $a$ and $b$, related to the total mass of the source $M=2a$ (from the $g_{tt}$ asymptotic behaviour) and to the magnetic dipole moment $\mu=2ab$ (from the $A_{\phi}$ asymptotic behaviour). The Bonnor line element (\ref{metric}) manifests relatively complicated singular behviour at $P=0$, $Q=0$, $Z=0$ and $Y=0$, nature of which is described in details in the paper \cite{Ward75}. However, here we are interested in the regular part of the spacetime $Z>0$ only, limited by the horizon at $r_{\rm h}=a+\sqrt{a^2+b^2}$. 

The Bonnor field can be interpreted as a superposition of two magnetic poles of strengths $-a$ and $a$, both with mass $a$, that are situated on the symmetry axis and separated by the coordinate distance $2b$ \cite{Ward75}.\footnote{In the case of $b=0$ the proper distance between the magnetic poles is nonzero and the geometry does not tend to the Schwarzschild one.} The problem how to keep this magnetic monopoles appart in an equilibrium was answered, e.g., by adding an external magnetic field into the solution \cite{Emparan00}. Such a configuration is analogous to two oppositely electrically charged Reissner--Nordstr\"{o}m masses, which can survive in equilibrium due to an induced electric field mechanism, as recently confirmed within the framework of an exact solution of the static Einstein–-Maxwell equations \cite{MankoArxiv,Manko07}. A sophisticated analysis of the electrostatic two-–body problem has been recently presented also in \cite{Bini07a,Bini07b}, completed by the investigation of electric field lines, exhibiting the Meisssner effect found in such system \cite{Bini08}.
     
Naturally, a number of more recent exact solutions have been discovered, which can represent a massive magnetic dipole in relativity (see, e.g., papers  \cite{Gutsunaev88,Gutsunaev89,Karas91,Manko92}). They differ from each other by the structure of the field near the central body, where singularities can occur. But despite a large number of them, Bonnor's solution \cite{Bonnor66} plays a prominent role thanks to its simplicity and explicit form that allows us to interpret the solution and describe its properties in great details. The metric (\ref{metric}) actually represents a magnetostatic limit of a more general exact solutions \cite{manko00,pachon06} suggested to describe the exterior field of a rotating neutron star. In the case of Bonnor's solution, the rotation is not considered and the value of a quadrupole mass moment is fixed by values of the mass and dipole parameters $a$ and $b$

\begin{eqnarray}
\label{Quadrupole}
\mathcal{Q}&=&\frac{\mu^2}{M}-\frac{1}{4}M^3=2a(b^2-a^2),
\end{eqnarray}
vanishing and changing its sign for $|b|=a$. These features support an exclusion of the pure Bonnor spacetime from astrophysically realistic ones (which are not static, instead they are dragged into a rotation by spinning source, often also keeping the quadrupole moment as another independent parameter). Thus, the investigation of particle motion in the Bonnor's static magnetic dipole background presented here has mainly a theoretical importance; for example, it allows us to study the onset of chaos in General Relativity. From the theoretical point of view, however, it is interesting to point out that the frame-dragging, being typically related with  spacetimes of spinning sources, can be also found in the vicinity of static source endowed with an electromagnetic structure \cite{Bonnor91,Herrera06}. In more details, as shown in the paper \cite{Herrera06} studying a rotating magnetic dipole bearing an electric charge, the vorticity tensor is nonvanishing and the frame-dragging occurs even in the special case when the angular momentum of the source is zero. This is due to the presence of the magnetic dipole moment and the additional electric charge of the source, which is, however, not the case of the Bonnor's massive dipole studied here.  

According to the standard approach to investigation of particle dynamics, we construct the super-Hamiltonian \cite{mtw}
\begin{equation}
\label{hamiltonian}
\mathcal{H}=\textstyle{\frac{1}{2}}g^{\mu\nu}(\pi_{\mu}-\tilde{q}A_{\mu})(\pi_{\nu}-\tilde{q}A_{\nu}),
\end{equation}
where $\tilde{q}$ is the charge of particle, $\pi_{\mu}$ is the generalized (canonical) momentum, $g_{\mu\nu}$ is the
metric tensor (\ref{metric}) and $A_{\mu}$ denotes the electromagnetic vector potential (\ref{EMpot}). The latter is related to the electromagnetic
tensor by $F_{\mu\nu}=A_{\nu,\mu}-A_{\mu,\nu}$. 
The related Hamiltonian equations which we integrate to obtain general trajectories of particles are given as
\begin{equation}
\label{hkr}
\frac{{\rm d}x^{\mu}}{{\rm d}\lambda}\equiv p^{\mu}=
\frac{\partial \mathcal{H}}{\partial \pi_{\mu}},
\quad 
\frac{d\pi_{\mu}}{d\lambda}=-\frac{\partial\mathcal{H}}{\partial x^{\mu}},
\end{equation}
where $\lambda=\tau/m$ is the affine parameter, $\tau$ denotes the
proper time, $m$ stands for the rest mass of the particle and $p^{\mu}$ is the standard kinematical four-momentum for
which the first equation reads $p^{\mu}=\pi^{\mu}-\tilde{q}A^{\mu}$.

The Bonnor's solution obeys two obvious symmetries defined by Killing vectors fields $\eta^{\mu}=\delta^{\mu}_t$ and $\xi^{\mu}=\delta^{\mu}_{\phi}$, which ensure the existence of two related constants of motion, quaranteed also by second Hamiltonian equation. These are the specific (per unit mass) energy and the specific angular momentum 
\begin{eqnarray}
\label{Momenta}
E&=&-u_t- q A_t=\left(\frac{P}{Y}\right)^2 \frac{dt}{d\tau},\\
L&=&u_{\phi}+q A_{\phi}=Z\sin^2{\theta} \left(\frac{Y}{P}\right)^2 \frac{d\phi}{d\tau} +q
A_{\phi},
\end{eqnarray}
where $q=\tilde{q}/m$ is the specific charge of particle.

The motion in the meridian plane can be described very well by the equation 
\begin{eqnarray}
E^2 =\frac{P^4}{Q^3Z}\ \left[ \left( \frac{dr}{d\tau} \right)^2 +Z\left( 
\frac{d\theta}{d\tau} \right)^2 \right]+V_{\mathrm{eff}}^2
\end{eqnarray}
with the effective potential given by the relation
\begin{equation}
V_{\mathrm{eff}}^2=\frac{P^2}{Y^2} \left[1 + \frac{P^2}{Y^2Z\sin^2{\theta}}
\left( L-q\mu\frac{r\sin^2{\theta}}{P} \right)^2 \right].
\end{equation}
Because of the properties of the metric (squared $b$ parameter) and the effective potential ($b$ parameter in product with $q$, squared $L-qA_{\phi}$ term), the effective potential is invariant under combinations of simultaneous sign reversals maintaining ${\rm sgn}(L q)$,  ${\rm sgn}(L b)$ or ${\rm sgn}(q b)$.  Since the metric is static and axially symmetric, we expect that closed, stable planar orbits should be possible in the
equatorial plane ($\theta=\pi/2$). The self-consistent magnetic field obeys the same symmetries in this solution, thus we can also expect
the possibility of the off-equatorial azimuthal circulation of electrically charged matter when the Lorentz force is in equilibrium
with gravitational and centrifugal forces. 

The circular orbits of constant radius and latitude (circular halo orbits) are determined by stationary points of the effective potential, i.e., by the conditions
\begin{eqnarray}
\label{stationary}
\partial_{r} V_{\rm eff}=0,\quad \partial_{\theta} V_{\rm eff}=0.
\end{eqnarray}
Being interested in the stability of the motion against the perturbations, we have to construct the Hessian matrix
\begin{eqnarray}
H =
\left( 
\begin{array}{cc}
\partial^2_r V_{\rm eff}&\partial_{r}\partial_{\theta} V_{\rm eff}\\ 
\partial_{\theta}\partial_{r} V_{\rm eff}&\partial^2_{\theta} V_{\rm eff}
\end{array}
\right).
\end{eqnarray}
Circular orbits at $r=r_{\rm c}$ and $\theta=\theta_{\rm c}$ stable against all the perturbations (minima of $V_{\rm eff}$) must satisfy the conditions 
\begin{eqnarray}
\label{stability1}
\partial^2_r V_{\rm eff}|_{r=r_{\rm c},\theta=\theta_{\rm c}}>0,\quad
\label{stabilty2}
\det H|_{r=r_{\rm c},\theta=\theta_{\rm c}}>0.
\end{eqnarray}
In particular, the stability of circular orbits against radial perturbations (important mainly in the case of equatorial motion) is guaranteed by the first from conditions (\ref{stability1}). On the other hand, fully unstable circular orbits (maxima of $V_{\rm eff}$) satisfy the conditions
\begin{eqnarray}
\label{unstability1}
\partial^2_r V_{\rm eff}|_{r=r_{\rm c},\theta=\theta_{\rm c}}<0,\quad
\label{unstability2}
\det H|_{r=r_{\rm c},\theta=\theta_{\rm c}}>0.
\end{eqnarray}
Investigating the circular motion, it is also useful to monitor the azimuthal component of the linear velocity 
\begin{eqnarray}
v\equiv \hat{v}_{\phi}=\frac{\hat{u}^{\phi}} {\hat{u}^t}=\frac{\sqrt{-g_{tt}}(L-qA_{\phi})}{E\sqrt{g_{\phi\phi}}}.
\end{eqnarray}

%%%%%%%%%%%%%%%%%%%%%%%%%%%%%%%%%%%%%%%%%%%%%%%%%%%%%%%%%%%%%%%%%%%%%%%%%%%%%%%%%%%%%%%%%%%%%%5

\section{Circular equatorial motion}
The equatorial circular motion of charged particle at radius $r$ is characterized by the specific angular momenta 
\begin{eqnarray}
\label{Lcpm}
L_{\rm c\pm}&=&\frac{q a b(2F+rZ)\pm r Z\sqrt{(2aF+(q a b)^2)}}{(r-2a)F},
\end{eqnarray}
following from the first of conditions (\ref{stationary}) or, alternatively, by the related azimuthal component of the linear velocity
\begin{eqnarray}
\label{vcpm}
v_{\rm c\pm}&=&v(L=L_{\rm c\pm},E=V_{\rm eff}).
\end{eqnarray}

In the special case of uncharged ($q=0$) particles, the reality condition $F=4ab^2+r(r-5a)(r-2a)\geq 0$ in relation (\ref{Lcpm}) restricts the existence of equatorial circular orbits by the photon circular orbits at radii given by the solutions of $F=0$, in the $(r\times b)$ plane determined by the curve defined by the functions
\begin{eqnarray}
b_{\rm ph}=\pm\sqrt{\frac{r(5a-r)(r-2a)}{4a}}.
\end{eqnarray}
Extrema of these functions are located at $r\doteq 3.786\,a$, taking values $b\doteq \pm 1.433\,a$. In the case of $b=0$, the inner photon orbit coalesces with the horizon at $r_{\rm h}=2\,a$, the outer one is located at $r=5\,a$.  
As it follows from the investigation of conditions (\ref{stability1}) with $\theta_{\rm c}=\pi/2$, in the $(r\times b)$ plane, the \mbox{$r$-region} of fully stable orbits and orbits unstable against radial perturbations are separated by the curve defined by the functions 
\begin{eqnarray}
\label{ms}
b^0_{\rm ms}&=&\pm\Bigg{(}\frac{\pm r(r-2a)\sqrt{(3r^2-10ar+6a^2)(3r^2+22ar-26a^2)}}{32a^2}-\\\nonumber
&&-\frac{3r(r^3-18a^2r+28a^3)}{32a^2}\Bigg{)}^{1/2},
\end{eqnarray}
maximum and minimum of which are located at $r\doteq 7.236\,a$ and take values $b=\pm 2.115\,a$. The curves $b^0_{\rm ms}$ and $b_{\rm ph}$ interstect each other just in the extrema of $b_{\rm ph}$. 
Fully stable orbits and orbits stable against radial perturbations (but not fully stable) are separated by the curve defined as
\begin{eqnarray}
\label{mcs}
b^0_{\rm mcs}=\pm r\sqrt{\frac{r-2a}{3r-2a}}.
\end{eqnarray}

\begin{figure}[tbh!]
\centering
\includegraphics[width=0.95\hsize]{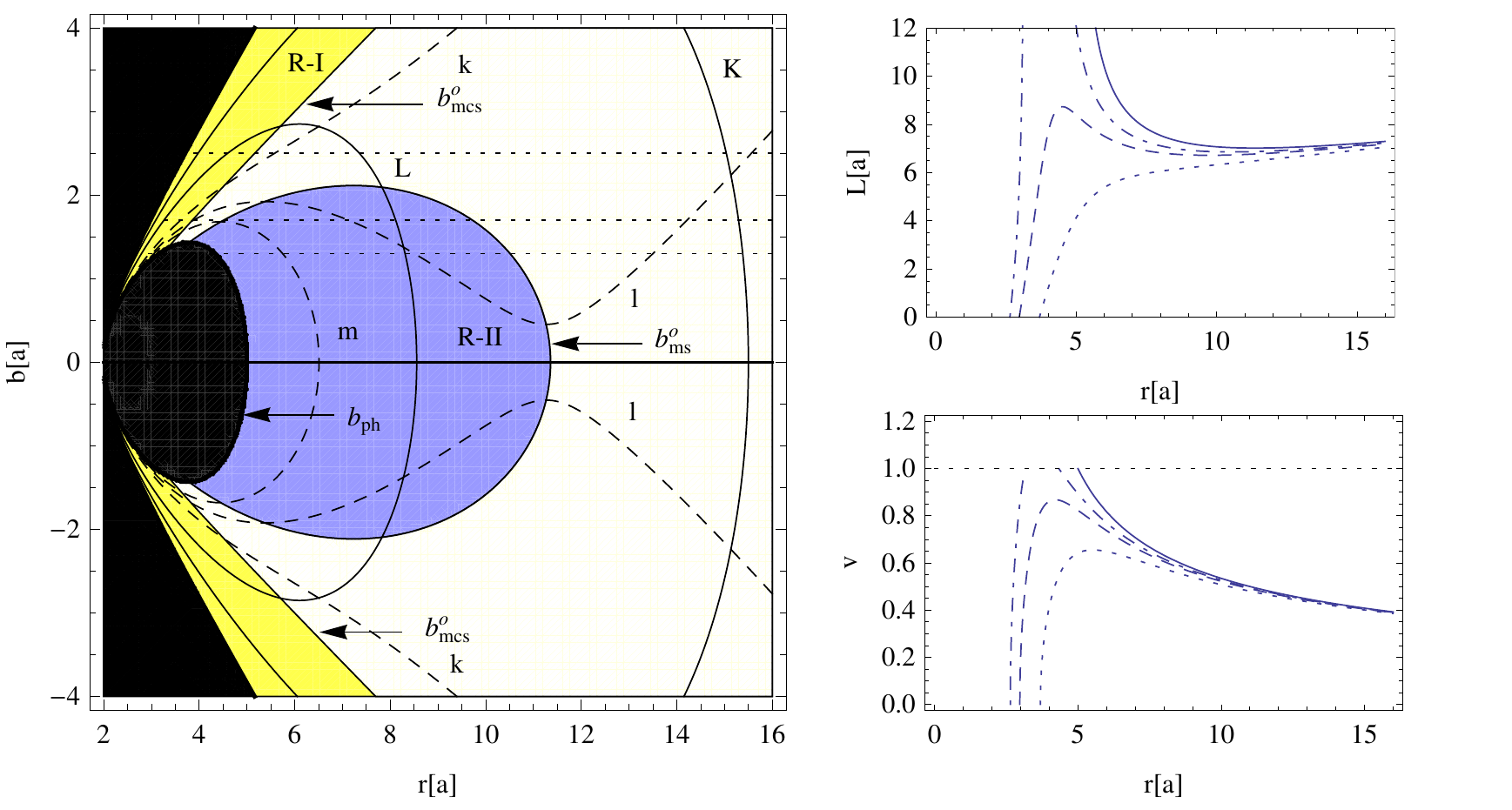}
\caption{{\bf Left:} Regions of stability for circular orbits of uncharged particles moving in the equatorial plane with the specific angular momentum profile given by $L^0_{\rm c+}$: region of orbits stable against all the perturbations (white), region of orbits stable against the radial perturbations but not fully stable (yellow R-I) limited by the curve $b^0_{\rm mcs}$ and region of orbits unstable against the radial perturbations (blue R-II) limited by the curve $b^0_{\rm ms}$. The circular orbits do not exist in the oval-like black region limited by the curve $b_{\rm ph}$ denoting positions of circular photon orbits; the triangle-like black regions represent the regions under the horizon $Z=0$. The dashed (k $\equiv5$, l $\equiv7$, m $\equiv9$) and solid (K $\equiv0.4$, L $\equiv0.6$) curves denote contours of specific angular momentum $L^0_{\rm c+}$ (in units of $a$) and related velocity $v^0_{\rm c+}$. The horizontal dotted lines and their intersections with the stability regions and limiting curves suggest the behaviour of $L^0_{\rm c+}$ and $v^0_{\rm c+}$ as shown in the right panels of this figure. {\bf Right:} Specific angular momentum $L^0_{\rm c+}$ and related velocity $v^0_{\rm c+}$ profiles for $b=0$ (solid), $b=1.3\,a$ (dot-dashed), $b=1.7\,a$ (dashed) and $b=2.5\,a$ (dotted). The angular momentum diverges at the radii of photon orbits, where the related velocity reaches the speed of light, and vanishes at the radius of horizon as well as the velocity.} 
\label{Fig:1}
\end{figure}

We present stability regions and the introduced functions in figure \ref{Fig:1} along with the typical behaviour of the specific angular momentum $L^0_{\rm c+}\equiv L_{\rm c+}(q=0)$ and velocity $v^0_{\rm c+}\equiv v_{\rm c+}(q=0)$, plotted for different values of $b$ from intervals defined by the maxima of $b_{\rm ph}$ and $b^0_{\rm ms}$.
Note that positions of minima of $L^0_{\rm c+}$ separate the $r$-regions of stability; the descending parts correspond to $r$-stable orbits and ascending parts to $r$-unstable orbits. For $b=0$, we find the stable orbits limited from below by $r_{\rm ms}\doteq 11.359\,a$. The specific angular momentum $L^0_{\rm c+}$ diverges at the radii of photon orbits (related velocity $v^0_{\rm c+}$ reaches the speed of light here) and vanish at the radius of horizon as well as the velocity.
Because of the absence of the electric charge, we have here two specific angular momentum profiles $L^0_{\rm c+}$ and $L^0_{\rm c-}$ (\ref{Lcpm}), differing only in the sign, and the effective potential resistant to a change of the sign of $L$. Consenquently, the stability regions illustrated in figure \ref{Fig:1}, calculated for $L=L^0_{\rm c+}$ profile, remains the same for $L^0_{\rm c-}$ profile, only the signs of specific angular momentum and velocity contours will be different.  

\begin{figure}[tbh!]
\centering
\includegraphics[scale=0.88, trim = 0mm 0mm 0mm 0mm, clip]{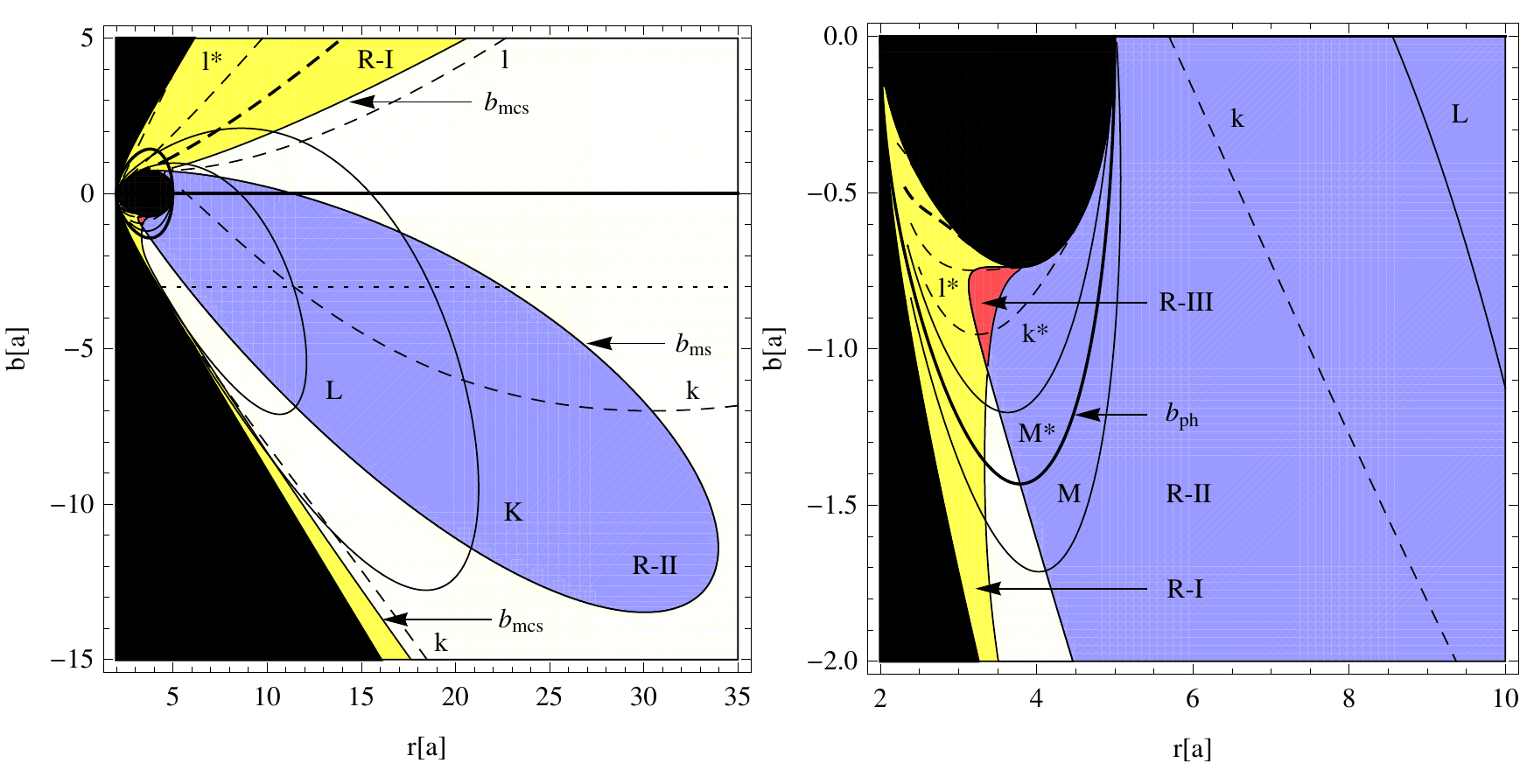}
\caption{Regions of stability in two different zooms for circular orbits of charged particles with $q=4.718$ moving in the equatorial plane with the specific angular momentum profile given by $L_{\rm c-}$: region of orbits stable against all the perturbations (white), region of orbits stable against the radial perturbations but not fully stable (yellow R-I) limited by the curve $b_{\rm mcs}$, region of orbits unstable against the radial perturbations (blue R-II) limited by the curve $b_{\rm ms}$ and region of orbits unstable against all the perturbations (red R-III). The circular orbits do not exist in the oval-like black region; the triangle-like black regions represent the regions under the horizon $Z=0$. The dashed (k $\equiv-12$, l $\equiv-4$, l$^*$ $\equiv4$, k$^*$ $\equiv12$)  and solid (K $\equiv-0.4$, L $\equiv-0.6$, M $\equiv-0.995$, M$^*$ $\equiv0.995$) curves denote contours of specific angular momentum $L_{\rm c-}$ (in units of $a$) and related linear velocity $v_{\rm c-}$. The thick solid oval-like curve $b_{\rm ph}$ denotes positions of circular photon orbits. The thick dashed curve denotes the zero specific angular momentum contour. The horizontal dotted line $b=-3\,a$ and its intersections with the stability regions and limiting curves suggest the behaviour of $L_{\rm c-}$ and $v_{\rm c-}$ as shown in figure \ref{Fig:3}.}
\label{Fig:2}
\end{figure}
\begin{figure}[tbh!]
\centering
\includegraphics[scale=0.79, trim = 0mm 0mm 0mm 0mm, clip]{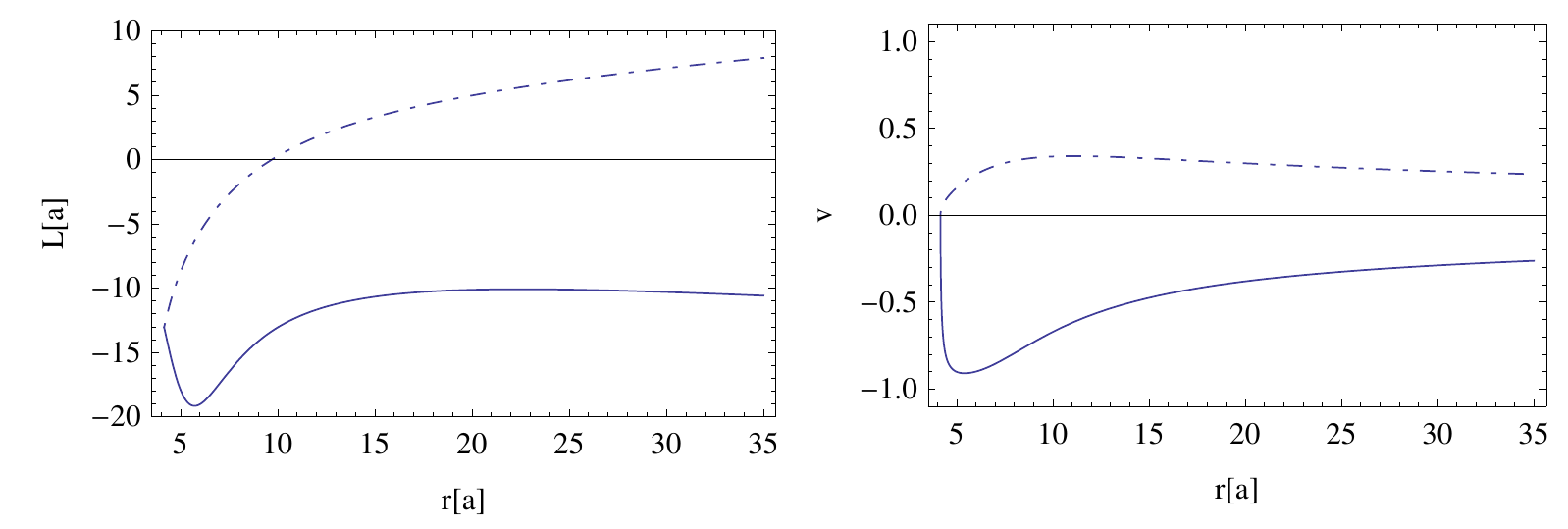}
\caption{Specific angular momenta $L_{\rm c-}$ (solid) and $L_{\rm c+}$ (dot-dashed) and related linear velocities $v_{\rm c-}$ (solid) and $v_{\rm c+}$ (dot-dashed) profiles for particles with specific charge $q=4.718$ moving in Bonnor spacetime with $b=-3\,a$. The angular momenta profiles coalesce with each other at the radius of horizon, where the related velocities vanish.}
\label{Fig:3}
\end{figure}
\begin{figure}[tbh!]
\centering
\includegraphics[scale=0.88, trim = 0mm 0mm 0mm 0mm, clip]{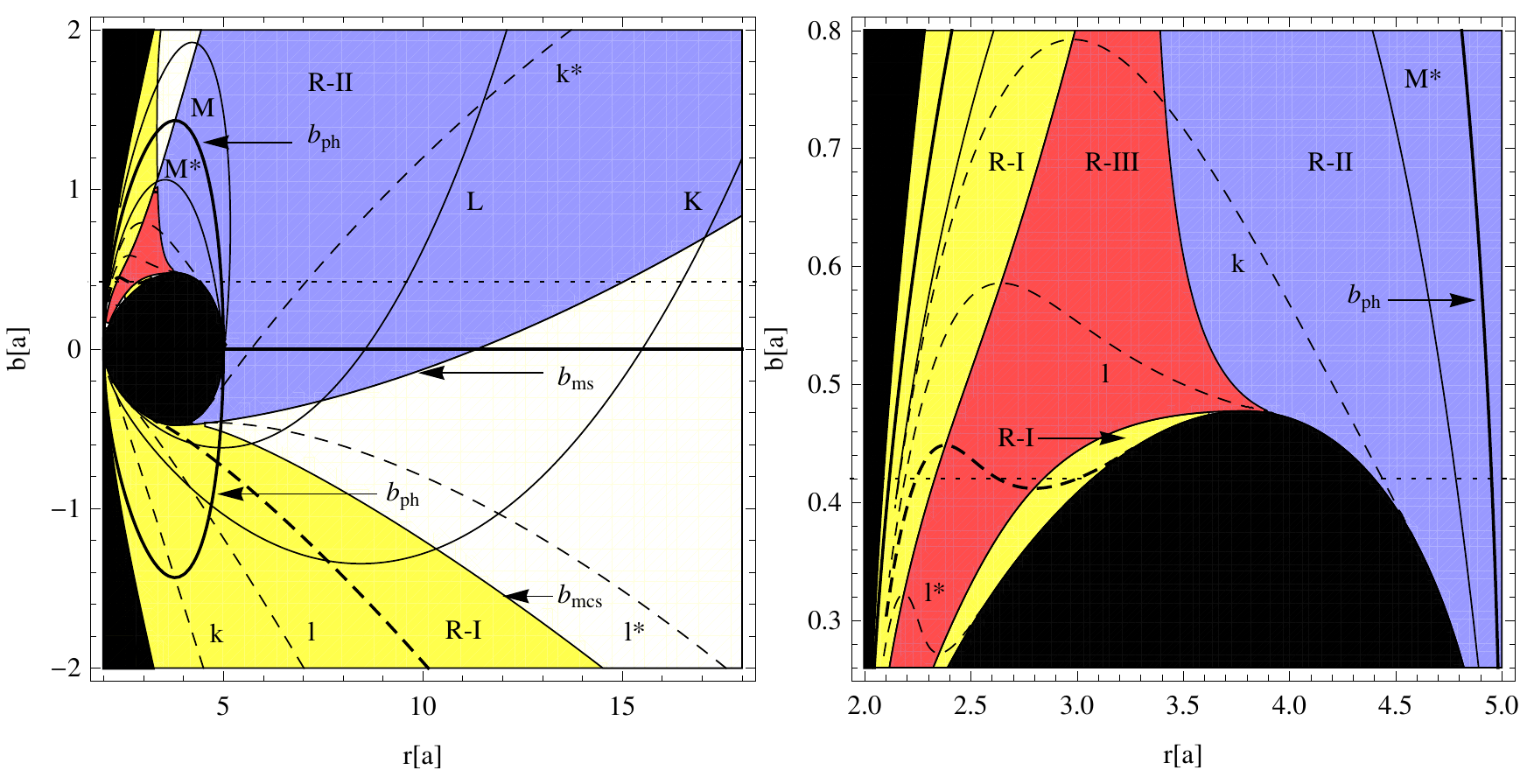}
\caption{Regions of stability in two different zooms for circular orbits of charged particles with $q=8$ moving in the equatorial plane with the specific angular momentum profile given by $L_{\rm c+}$: region of orbits stable against all the perturbations (white), region of orbits stable against the radial perturbations but not fully stable (yellow R-I) limited by the curve $b_{\rm mcs}$, region of orbits unstable against the radial perturbations (blue R-II) limited by the curve $b_{\rm ms}$ and region of orbits unstable against all the perturbations (red R-III). The circular   orbits do not exist in the oval-like black region; the triangle-like black regions represent the regions under the horizon $Z=0$. The dashed (k$\equiv-12$, l $\equiv-4$, l$^*$ $\equiv4$, k$^*$ $\equiv12$) and solid (K $\equiv0.4$, L $\equiv0.6$, M $\equiv0.995$, M$^*$ $\equiv-0.995$) curves denote contours of specific angular momentum $L_{\rm c+}$ (in units of $a$) and related linear velocity $v_{\rm c+}$. The thick solid oval-like curve $b_{\rm ph}$ denotes positions of circular photon orbits. The thick dashed curve denotes the zero specific angular momentum contour. The horizontal dotted line $b=0.42\,a$ and its intersections with the stability regions and limiting curves suggest the behaviour of $L_{\rm c+}$ and $v_{\rm c+}$ as shown in figure \ref{Fig:5}.} 
\label{Fig:4}
\end{figure}

In the general case of charged particles $q\neq 0$, we get the functions (\ref{ms}) and (\ref{mcs}) heavily dependent on the specific charge, i.e., we have $b_{\rm ms}\equiv b_{\rm ms}(r;a,q)$ and $b_{\rm mcs}\equiv b_{\rm mcs}(r;a,q)$. The relevant expressions, however, are rather long to be mentioned here and the analytical discusion is very complex. Nevertheless, in figure \ref{Fig:2}, we plot the representative behaviour of these functions for the fixed value $q=4.716$, giving us one of the most complex configurations of the stability regions.   
The stability regions shown in figure \ref{Fig:2} are constructed providing $L=L_{\rm c-}$ and, will change when $L=L_{\rm c+}$. But because of the symmetries of the effective potential, just in the way of inverting along the $b=0$ axis and changing the signs of specific angular momentum and velocity contours. In figure \ref{Fig:4}, we plot the stability regions from this class as well, i.e., corresponding to the $L=L_{\rm c+}$ root, and setting $q=8$. 
\begin{figure}[tbh!]
\centering
\includegraphics[scale=0.78, trim = 0mm 0mm 0mm 0mm, clip]{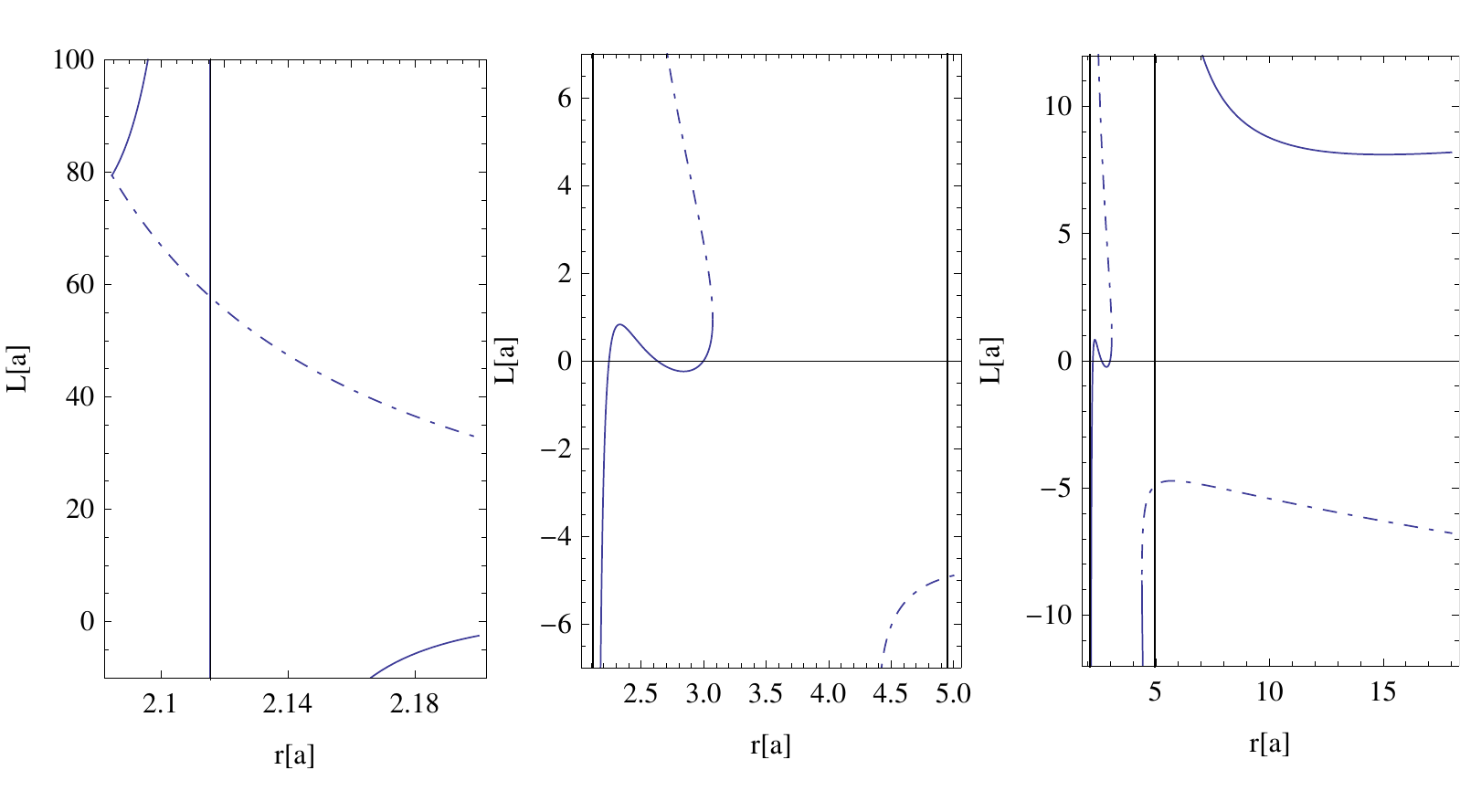}
\includegraphics[scale=0.78, trim = 0mm 0mm 0mm 0mm, clip]{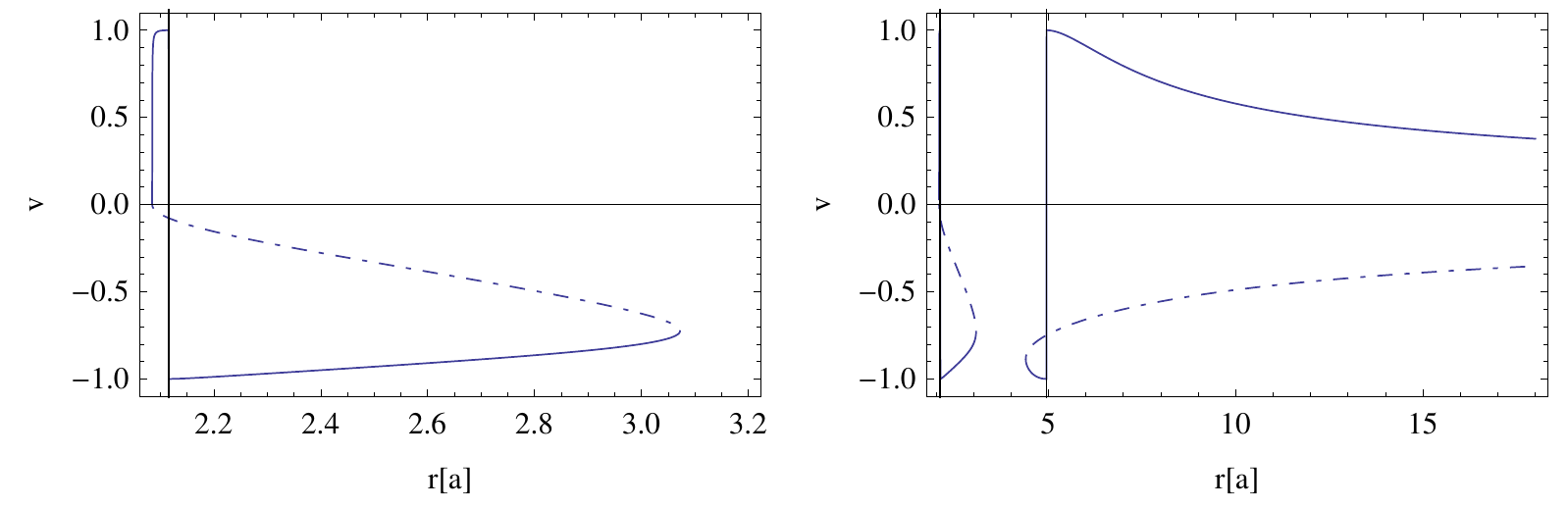}
\caption{Specific angular momenta $L_{\rm c+}$ (solid) and $L_{\rm c-}$ (dot-dashed) and related linear velocities $v_{\rm c+}$ (solid) and $v_{\rm c-}$ (dot-dashed) profiles for particles with specific charge $q=8$ moving in Bonnor spacetime with $b=0.42\,a$, plotted in different zooms. The $L_{\rm c+}$ profile diverges at the radii of photon orbits (vertical solid lines), where the related velocity reach the speed of light, and coalesce with $L_{\rm c-}$, among others, at the radius of horizon, where the related velocities vanish. The $L_{\rm c-}$ profile takes finite values at the radii of photon orbits, where the related velocity is sublight.} 
\label{Fig:5}
\end{figure}

Positions of minima of $L_{\rm c\pm}$ separate the $r$-regions of stability, the descending parts correspond to $r$-stable orbits and ascending parts to $r$-unstable orbits, as well as in the uncharged case. The situation here, however, is generally much more interesting. 
At first, note that thanks to the electric charge, we can find also fully unstable circular orbits, satisfying conditions (\ref{unstability1}). 
Moreover, we have two completely different specific angular momentum profiles $L_{\rm c+}$ and $L_{\rm c-}$ (\ref{Lcpm}) here. Thus, for instance, for a special combination of the spacetime parameter $b$ and specific charge $q$, the $L_{\rm c+}$ profile diverges at the radii of photon orbits (related $v_{\rm c+}$ reaches the speed of light here) and vanishes at the radius of horizon as well as $v_{\rm c+}$. We call the related circular orbits Larmor orbits. On the other hand, the $L_{\rm c-}$ profile takes finite values at the radii of photon orbits where the related $v_{\rm c-}$ are sublight. The related orbits are then anti-Larmor. Apparently, the circular orbits of charged particles can exist in the region between the photon orbits as well. Such a situation is well illustrated in figure \ref{Fig:5}.  

\section{Circular halo motion}
\begin{figure}[tbh!]
\centering
\includegraphics[scale=0.82, trim = 0mm 0mm 0mm 0mm, clip]{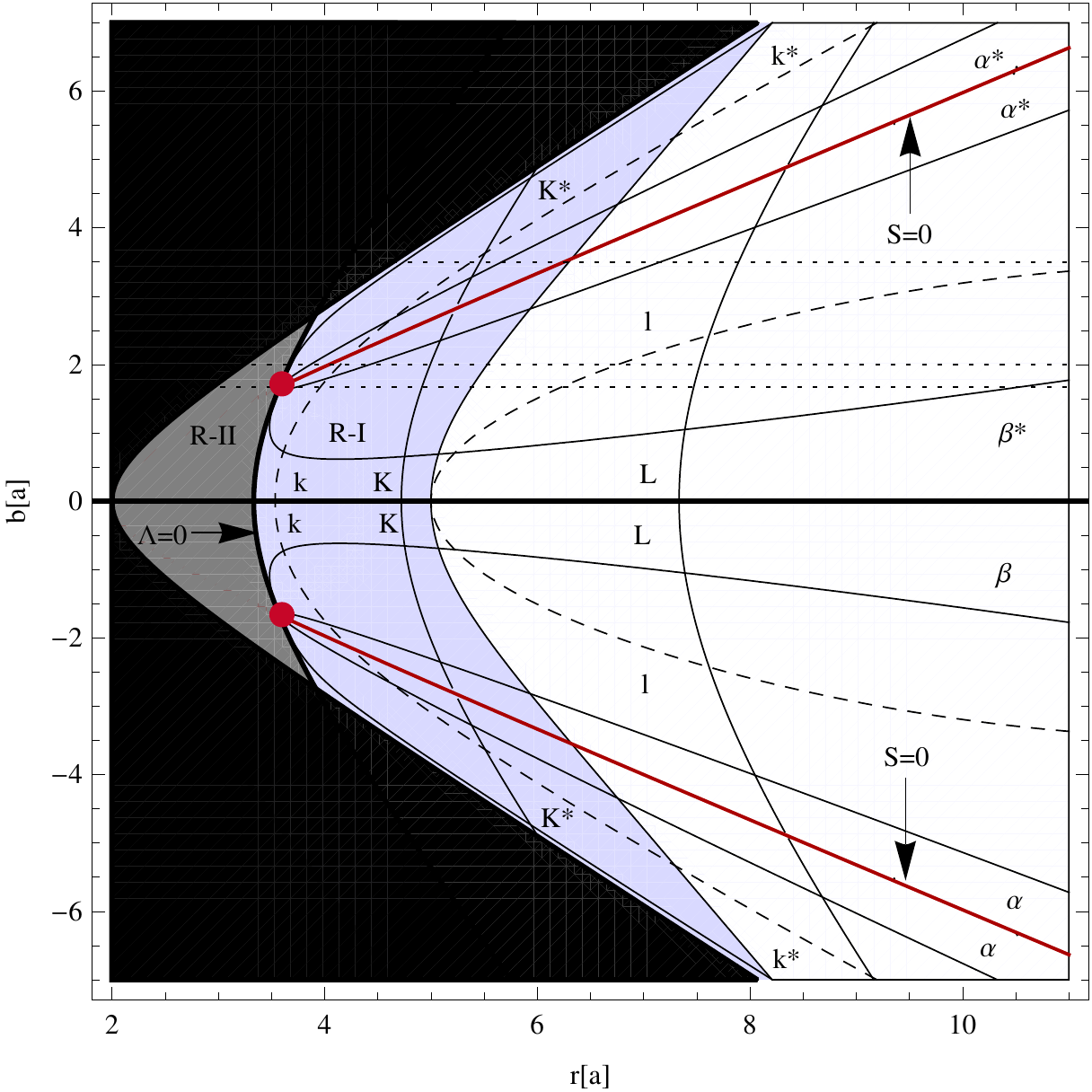}
\caption{Regions of stability for circular halo orbits of charged particles with the specific angular momenta and charges profiles given by $L_{\rm h+}$ and $q_{\rm h+}$: region of orbits stable against all the perturbations (white), region of orbits with an instability (blue R-I) and region with no circular halo orbits (gray R-II). The black regions denote the area under the horizon $Z=0$. The solid bow-like (more vertical) curves (K $\equiv-0.7$, L $\equiv-0.5$, L$^*$ $\equiv0.5$, K$^*$ $\equiv0.7$) denote contours of linear velocity $v_{\rm h+}$, the dashed curves (k $\equiv-6$, l $\equiv-2.5$, k$^*$ $\equiv6$) denote the contours of specific angular momentum $L_{\rm h+}$ (in units of $a$) and the solid (more horizontal) curves ($\alpha\equiv-0.7$, $\beta\equiv-7$ $\alpha^*\equiv0.7$, $\beta^*\equiv7$) denote the contours of specific charge $q_{\rm h+}$. The thick solid curve separating the regions R-I and R-II corresponds to part of the $\Lambda=0$ curve; the transversal thick red curves correspond to the $S=0$ curve, denoting positions of circular halo geodesics (dividing the positive and negative contours of $L_{\rm h+}$ and $v_{\rm h+}$); the positions of circular halo photon orbits correspond to their intersections (red bullets). The horizontal dotted lines $b=3.5\,a$, $b=2\,a$ and $b\doteq 1.668\,a$ denote the values for which the specific angular momenta, linear velocities and specific charges profiles are illustrated in figure \ref{Fig:7}.} 
\label{Fig:6}
\end{figure}

\begin{figure}[tbh!]
\centering
\includegraphics[scale=0.81, trim = 0mm 0mm 0mm 0mm, clip]{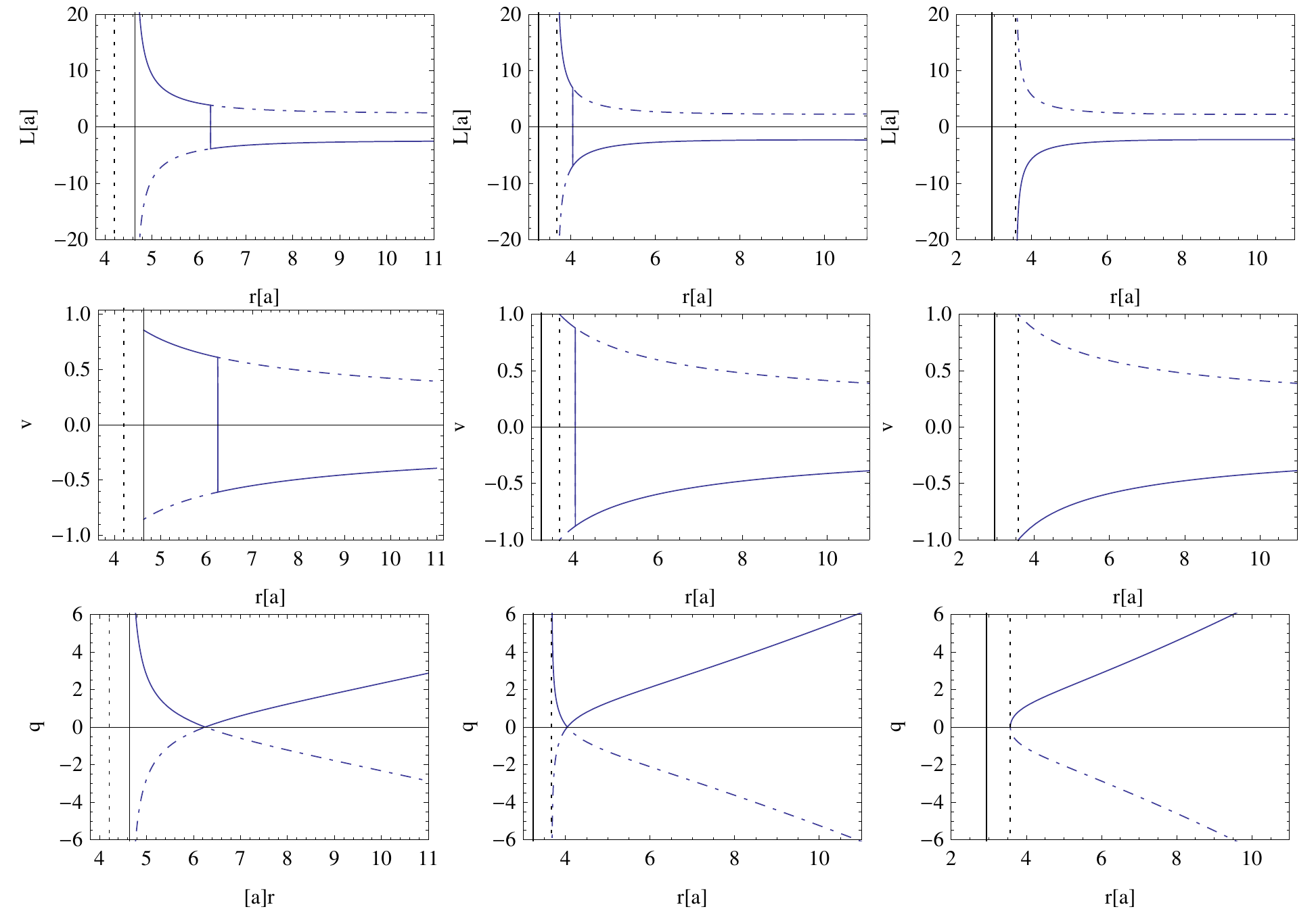}
\caption{Specific angular momenta $L_{\rm h+}$ (solid) and $L_{\rm h-}$ (dot-dashed), specific charges $q_{\rm h+}$ (solid) and $q_{\rm h-}$ and related linear velocities $v_{\rm h+}$ (solid) and $v_{\rm h-}$ for particles moving in Bonnor spacetimes with $b=3.5\,a$, $b=2\,a$ and $b\doteq 1.668\,a$. The angular momenta and charge profiles diverge at the radii of horizon (vertical solid line) given by $Z=0$ or at the radii given by $\Lambda=0$ (vertical dashed line), if they are above the horizon. Of course, at the radii given by $S=0$ (determining uncharged circular halo orbits) the charges vanishes. The velocities reach the speed of light at the radii given by $\Lambda=0$.} 
\label{Fig:7}
\end{figure}

Loci of circular halo orbits ($\theta \neq \pi/2$) determined by conditions (\ref{stationary}) are characterized by the specific angular momenta and charges
\begin{eqnarray}
\label{Lh}
L_{\rm h\pm}=\pm\frac{-Y \sin\theta(r^2+b^2)\sqrt{ra}}{\sqrt{Z\Lambda(r^2+b^2\cos^2\theta)}}\;{\rm sgn}(S),
\end{eqnarray}
\begin{eqnarray}
\label{qh}
q_{\rm h\pm}=\pm\frac{1}{2ab}\sqrt{\frac{aY^{2}S^{2}}{rZ\Lambda \sin ^{2}{\theta}(r^{2}+b^{2}\cos ^{2}{\theta })}}, 
\end{eqnarray}
where 
\begin{eqnarray}
S=r^{4}-2ar^{3}-3b^{2}r^{2}\sin ^{2}{\theta }+2ab^{2}r(2\sin ^{2}{\theta}-1)-b^{4}\cos ^{2}{\theta },
\end{eqnarray}
\begin{eqnarray}
\Lambda =3r^{4}-10ar^{3}-2b^{2}r^{2}\cos ^{2}{\theta }-6ab^{2}r\cos ^{2}{\theta}-b^{4}\cos ^{4}{\theta }.
\end{eqnarray}
The related linear velocity can be specified as
\begin{eqnarray}
v_{\rm h\pm}=v(L=L_{\rm h\pm},q=q_{\rm h\pm},E=V_{\rm eff}).
\end{eqnarray}
Apparently, the existence of circular halo orbits in the Bonnor spacetime in the region $Z>0$ is determined by the condition $\Lambda>0$.
Next, we are mainly interested in the stable motion, i.e., in the orbits which satisfy conditions (\ref{stability1}) for the local minima of the
potential. The regions of existence and stability of the circular halo orbits are illustrated for the representative latitude $\theta=\pi/3$ in figure \ref{Fig:6}, providing that $L=L_{\rm h+}$ and $q=q_{\rm h+}$. It is important to point out that the regions of existence and stability remain the same  after taking the other roots $L_{\rm h-}$ and $q_{\rm h-}$ into account for conditions (\ref{stability1}) analysis. Moreover, the figure appears qualitatively the same after changing $\theta$. We also plot behaviour of the angular momenta, related velocity and specific charges profiles in figure \ref{Fig:7}.  

\begin{figure}[tbh!]
\centering
\includegraphics[scale=0.79, trim = 0mm 0mm 0mm 0mm, clip]{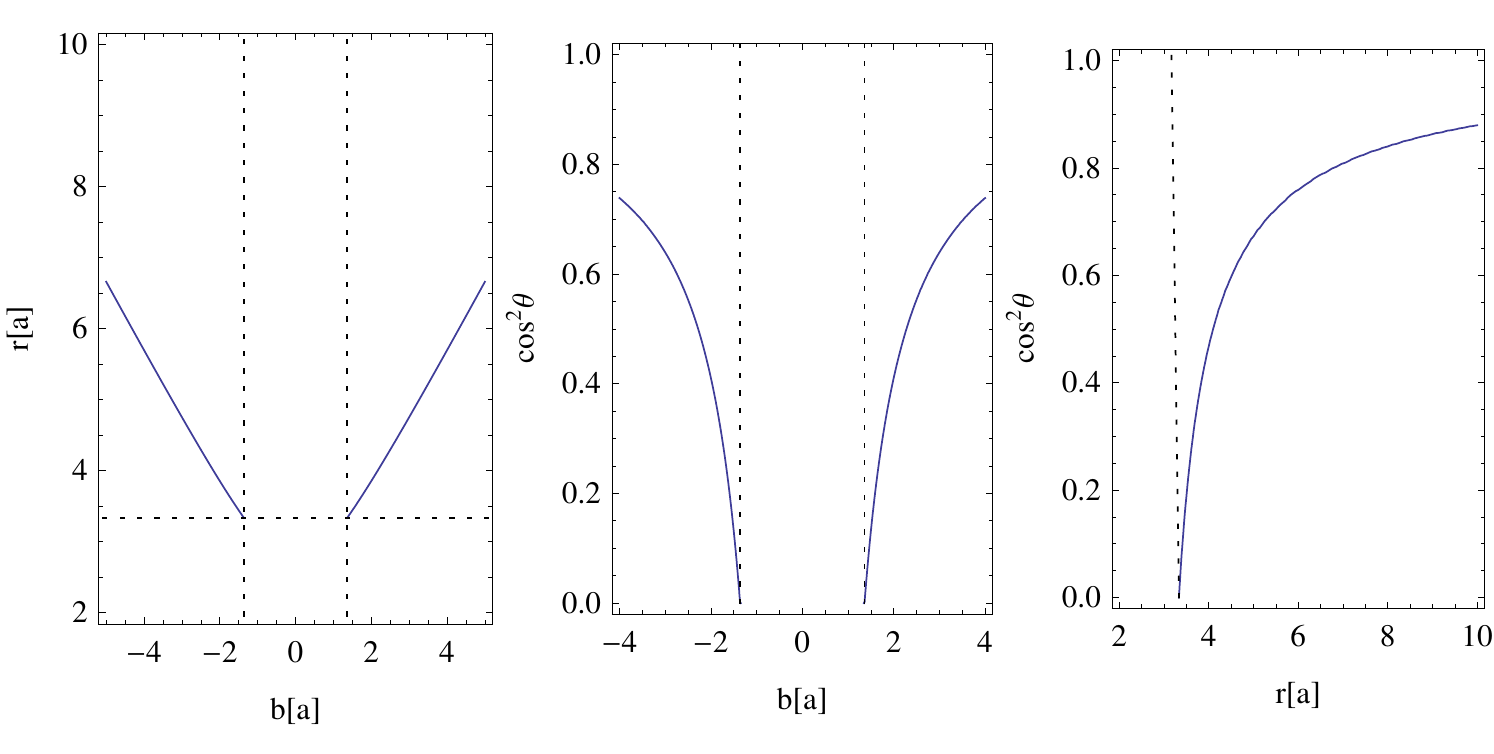}
\caption{Positions of circular halo photon orbits in Bonnor spacetimes. Orbits can occur only at radii $r>3.333$ and in spacetimes with $|b|>1.361$.} 
\label{Fig:8}
\end{figure}

Apparently, the special geometry of the Bonnor spacetime enables the existence of the circular halo orbits of uncharged particles (circular halo geodesics) as well. As it follows from relations $\partial_r V_{\rm eff}|_{q=0}=0$ and $\partial_{\theta} V_{\rm eff}|_{q=0}=0$ after eliminating $L$ or, directly from relation (\ref{qh}), their positions are determined by the condition $S=0$. The angular momentum of the uncharged particles moving along circular halo geodesics are determined by the relation $L^0_{\rm h\pm}=L_{\rm h\pm}/{\rm sgn}(S)$ (where the radii and latitudes are related by the formula $S=0$). The relation for $L_{\rm h\pm}$ diverges for $\Lambda=0$, therefore the condition $\Lambda=0$ together with $S=0$ enable us to determine positions of even circular halo photon orbits, as illustrated in figure $\ref{Fig:8}$. The reality conditions here restrict their existence to the regions $r>3.333$ and for spacetimes with $|b|>1.361$. 

\section{Regular and chaotic dynamics}
In this section, we perform a brief survey of dynamics of test particles moving within potential wells formed along both the equatorial and halo circular orbits. In order to do so, we apply several complementary methods of investigation of nonlinear dynamic systems. First of all, we construct Poincar\'{e} surfaces of section which give an overall perspective of the phase space dynamics on a given energy hypersurface (for a given values of system parameters). For the inspection of individual trajectories, however, we prefer to analyse their recurrence plots \cite{marwan07}, which proved to be very useful method in our previous work \cite{kopacek10}. Besides other properties of recurrence plots (RPs), we highlight their ability to clearly distinguish between chaotic and regular dynamics on a short time scale, thus reducing the integration time needed for the analysis. We also investigate intrinsic frequencies of the orbits employing the rotation number \cite{contopoulos02} that allows us to detect and locate resonances of the system. Recently, the methods of Poincar\'{e} surfaces of section and recurrence plots were very well applied, e.g., in
the papers \cite{Stuchlik12} and \cite{Semerak12}. 
\begin{figure}[htb]
\centering
\includegraphics[scale=0.46,trim=0mm 0mm 0mm 0mm,clip]{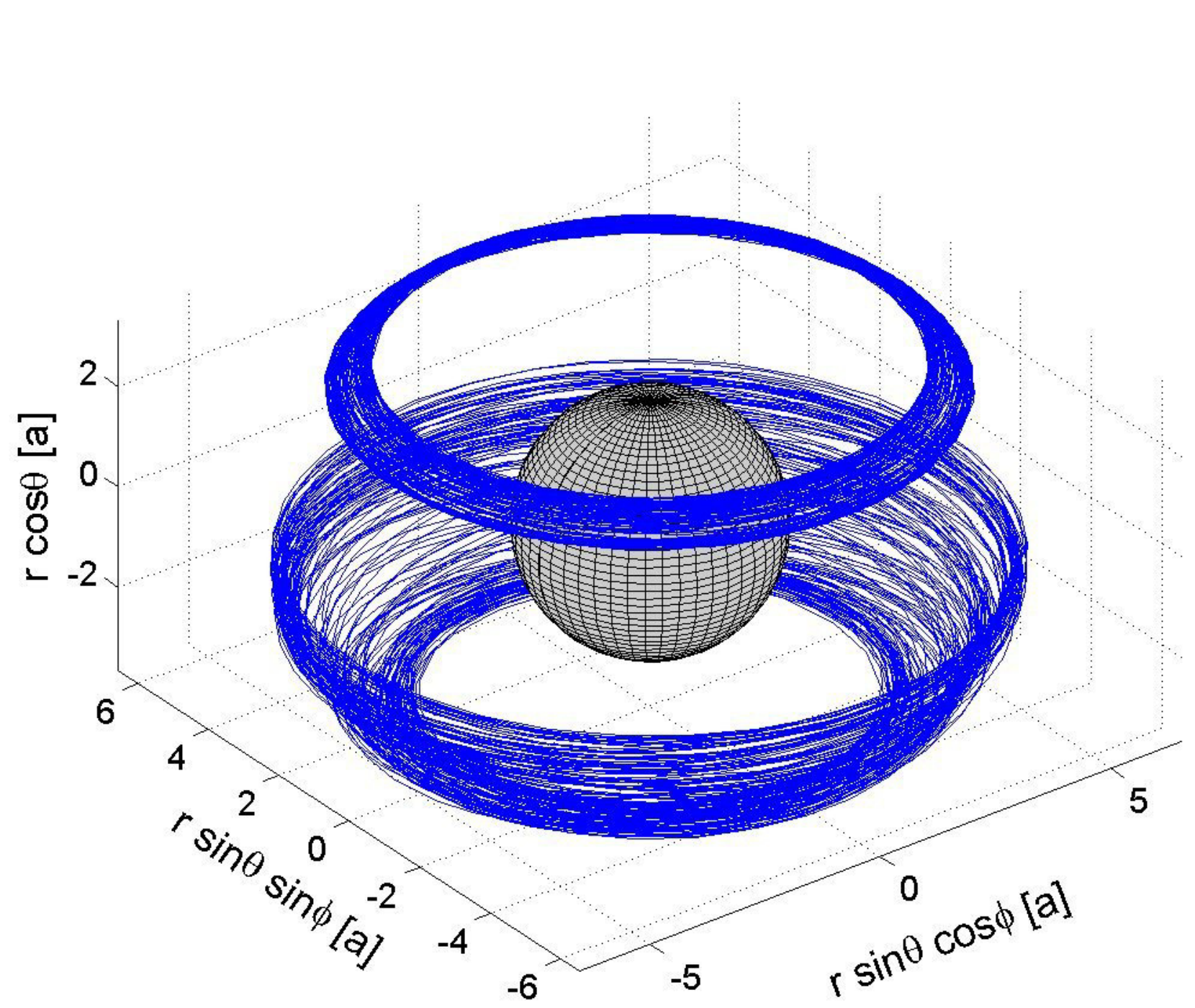}~~
\includegraphics[scale=0.39,trim=0mm 0mm 0mm 0mm,clip]{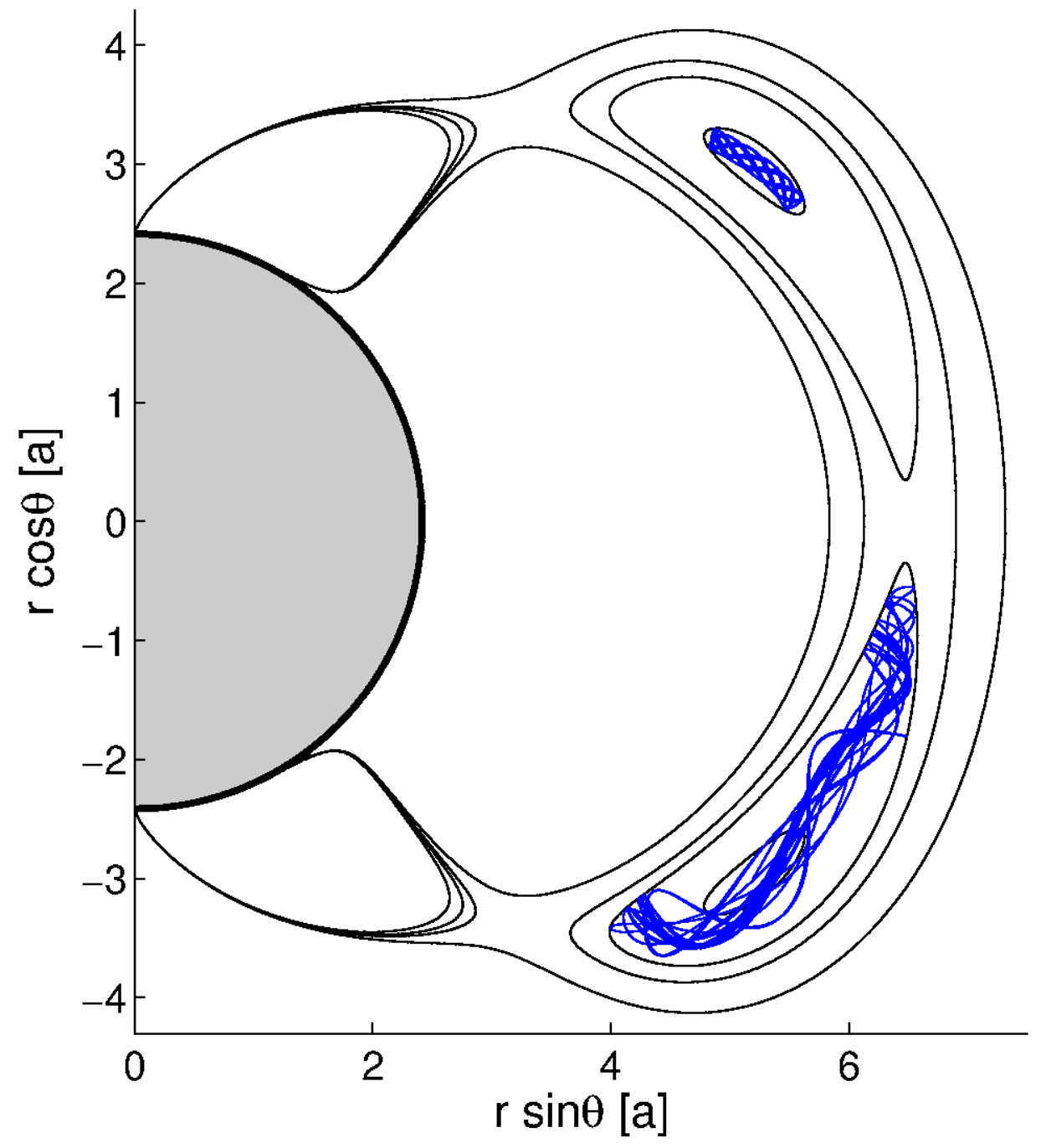}
\caption{Off-equatorial trajectories of charged test particle with $L=-2.356\,a$ and $q=5.581$ bounded in halo lobes of the Bonnor spacetime with $b=1\,a$. In the left panel we present a stereometric projection of two trajectories: the upper one with $E=0.8169$ shows ordered motion while with the higher energy the dynamics acquires properties of deterministic chaos (bottom trajectory with $E=0.8182$). We plot the poloidal $(r,\theta)$ projection of these trajectories along with several iso-contours of the effective potential in the right panel. Both particles were launched at $r(0)=6$, $\theta(0)=\pi/3$ with $u^r(0)=0$. Grey color marks $r=r_{\rm h}$ surface in both plots.}
\label{Fig:9}
\end{figure}

We are dealing with the autonomous Hamiltonian system (\ref{hkr}) describing the motion of electrically charged test particle around a massive center with dipole magnetic field described by the Bonnor metric (\ref{metric}). Due to the axial symmetry and stationarity of such a setup, we have two obvious constants of motion $E$ and $L$, related to cyclic coordinates $t$ and $\phi$, respectively. Besides that, the value of super-Hamiltonian (\ref{hamiltonian}) is also conserved, which corresponds to the conservation of the particle rest mass. The motion of a given particle thus generally occurs on the 3-dimensional hypersurface in the phase space. If there is, in addition, another independent integral of motion, analogical to the Carter's fourth constant in the Kerr-Newman spacetime \cite{carter68,mtw}, the system is completely integrable. In such a system, we could find no traces of the deterministic chaos at all and, each phase space trajectory would be bound to the 2-dimensio
 nal surface of 3-torus. On the other hand, if there is no extra constant of motion, the system is nonintegrable, which means that regular and chaotic orbits coexist in the phase space. In general, the amount of chaos depends on the strength of nonintegrable perturbation of the originally integrable system. In our case, however, such distinction between integrable and nonintegrable parts of the Hamiltonian is not obvious, because even if we set $b=0$, we do not obtain spherically symmetric Schwarzschild spacetime in which the motion of test particle is known to be integrable. Moreover, we have two types of possible perturbations: the magnetic field parametrized by $b$, which itself affects the motion of uncharged particles, and the electromagnetic force (proportional to charge $q$) exerted on the particle in the general case $q\neq0$, $b\neq0$.

We construct Poincar\'{e} surfaces of section from the intersection points of trajectories with the given $\theta=\theta_{\rm sec}=const$ plane. Instead of plotting the points in pair of canonical variables $(r,\pi_r)$, we translate the momenta into more intuitive quantity $u^r$ and, plot the $(r,u^r)$ pairs instead, which does not affect the interpretation of Poincar\'{e} sections. Given dimensionality of our system allows us to distinguish between the regular and chaotic orbits easily. The regular orbit bound to the surface of 3-dimensional torus in the phase space forms a narrow curve (with zero width, in principle), while the chaotic trajectory fills a nonzero surface on the section. In the integrable system, one finds regular orbits only. Both types of dynamics, however, are generally present in the nonintegrable case. 

Recurrence plots (RPs) are introduced as a tool of visualizing recurrences of a trajectory $\vec{x}(t)$
in the phase space \cite{eckmann87}. The construction of a RP is straightforward regardless of the dimension of
the phase space. First, we evaluate the binary values of the recurrence matrix $\mathbf{R}_{ij}$, which can be formally expressed as
\begin{equation}
\label{rpdef}
\mathbf{R}_{ij}(\varepsilon)=\Theta(\varepsilon-||\vec{x}(i)-\vec{x}(j)||),\;\;\;
i,j=1,...,N,
\end{equation}
where $\varepsilon$ is a pre-defined threshold parameter, $\Theta$ the Heaviside step function, and $N$ specifies the sampling frequency. The
sampling frequency is applied to the time segment of the trajectory $\vec{x}(t)$ under examination. There is, however, no unique prescription for the appropriate definition of the phase space norm $||\;.\;||$ and, simple $L^p$ norms are often applied. Particularly, $L^1$, $L^2$ (Euclidean norm) or $L^{\infty}$ (maximum norm) are the typical choice. Nevertheless, we employ a slightly different method (see \cite{kopacek10} for details on how we measure distances in the phase space for the detection of recurrences). The recurrence plot is obtained by direct visualizing of the matrix $\mathbf{R}_{ij}$ -- black dot is assigned in place of unity values, white dot reflects zero values of the matrix elements. The interpretation of a resulting plot along with its possible quantification is described in details in \cite{marwan07}.  

For the analysis of underlying frequencies of the orbit, we use the rotation number $\nu$. Its evaulation requires, at first, to localize the central fixed point ${u_0}$ in the Poincar\'{e} surface of section which is found in the common center of the primary tori. Being equipped with the set of section points of a given trajectory, we calculate the angle $\vartheta_i\equiv \angle (l_i,\,l_{i+1})$ between the radius vectors (defined with respect to ${u_0}$) of each pair of succeeding section points $u_i$ and $u_{i+1}$. The rotation number $\nu$ is defined as 
\begin{equation}
\label{rotationn}
\nu=\lim_{N\to\infty}\frac{1}{2\pi N}\sum_{i=1}^N \vartheta_i.
\end{equation}
We actually compute the mean of the finite number of values instead of the limit. 

In the following, we plot the rotation number as a function of the initial value of the particle radial coordinate, which reflects its position with respect to the central periodic orbit $u_0$. In such a plot we generally expect to find: i) smooth non-constant segments corresponding to regular orbits bound to primary tori with irrational frequency ratios (quasiperiodic orbits), ii) plateaus of various width reflecting resonant islands of various prominence, iii) `stochastic' segments resulting from chaotic zones, since the rotation number is not well defined for the chaotic orbits (the limit does not converge). Secondary islands of the stability are predicted by Poincar\'e-Birkhoff theorem to appear under a slight perturbation of the integrable system. They form chains of multiplicity given by the denominator of the rotation number, which is rational in this case. Each island consists of elliptic fixed point nested in the second order tori. The chain is completed by the same 
 number of unstable (hyperbolic) fixed points found between each pair of islands. Unstable fixed point appears as an inflection occurring at the same level of $\nu$ as the corresponding plateau in the $\nu(r)$ plot. However, resonant chains of islands of a single multiplicity also appear in some integrable systems. On the other hand, the presence of at least two types of islands (with different multiplicities) is regarded as strong indication of non-integrability, since no analytic integrable system with this property is known so far \cite{contopoulos02}.  

\section{Motion in equatorial lobes}
\begin{figure}[hp!]
\centering
\includegraphics[scale=0.44,trim=0mm 0mm 0mm 0mm,clip]{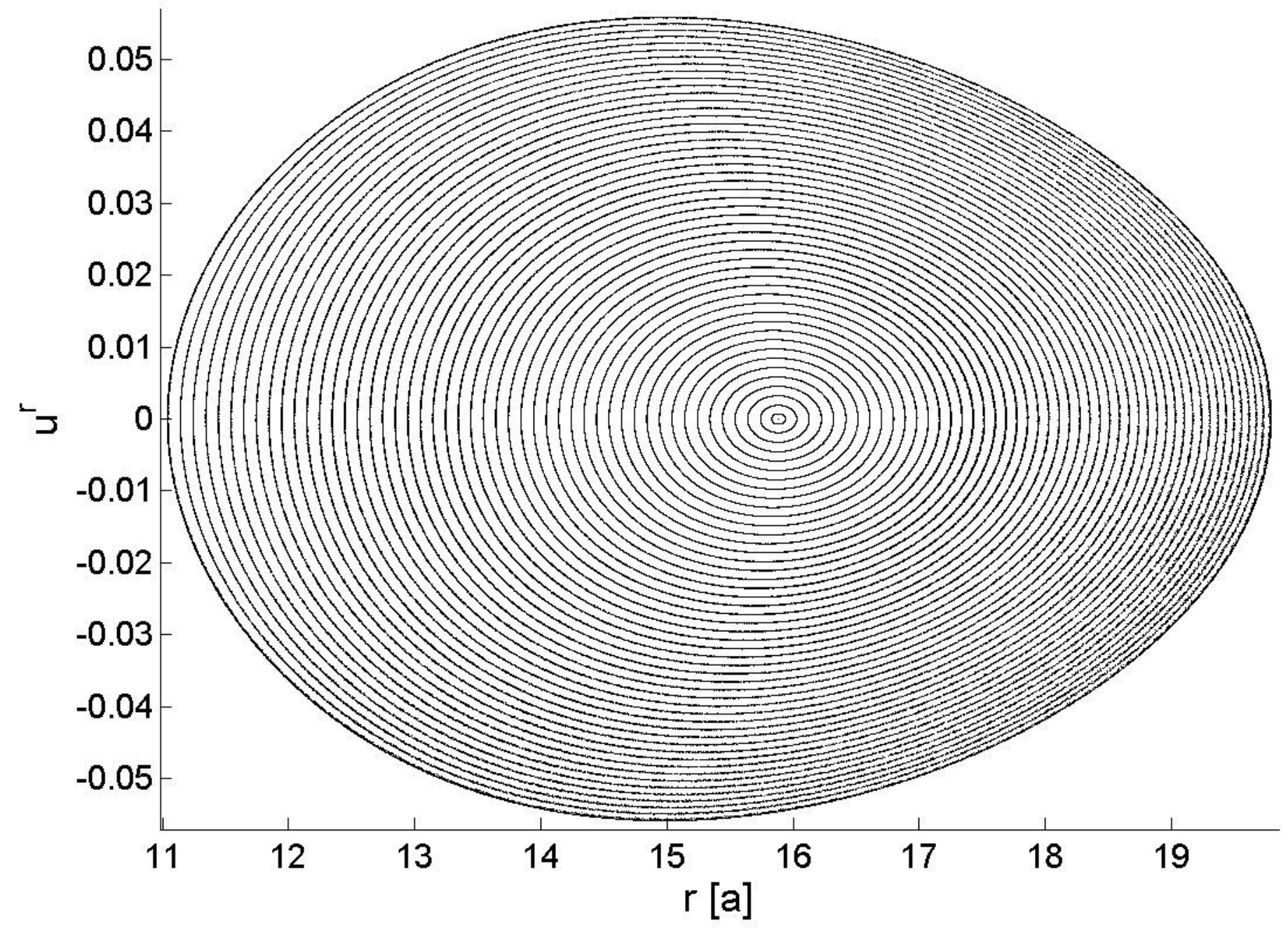}
\includegraphics[scale=0.41,trim=0mm 0mm 0mm 0mm,clip]{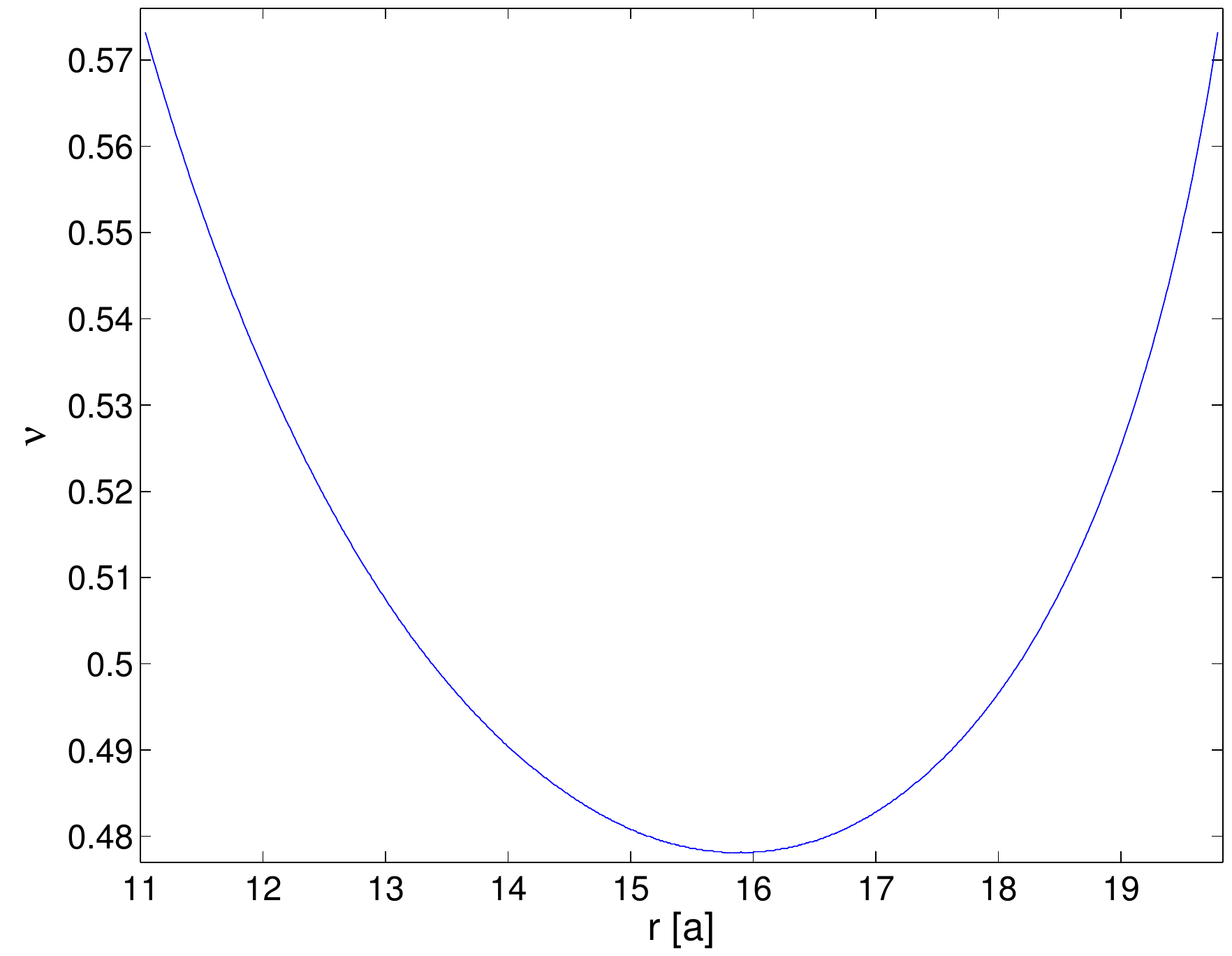}\\
\includegraphics[scale=0.43,trim=0mm 0mm 0mm 0mm,clip]{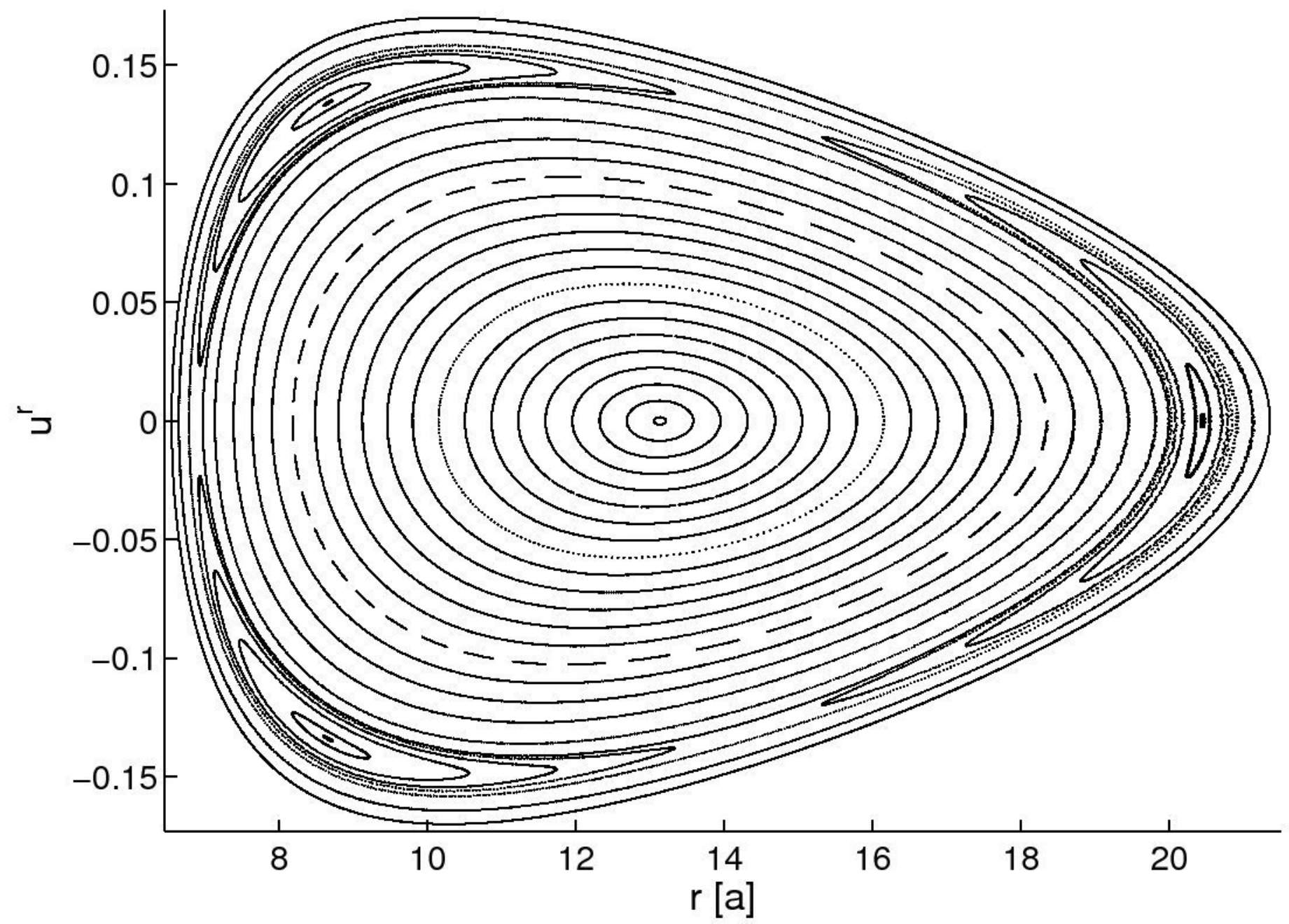}
\includegraphics[scale=0.44,trim=0mm 0mm 0mm 0mm,clip]{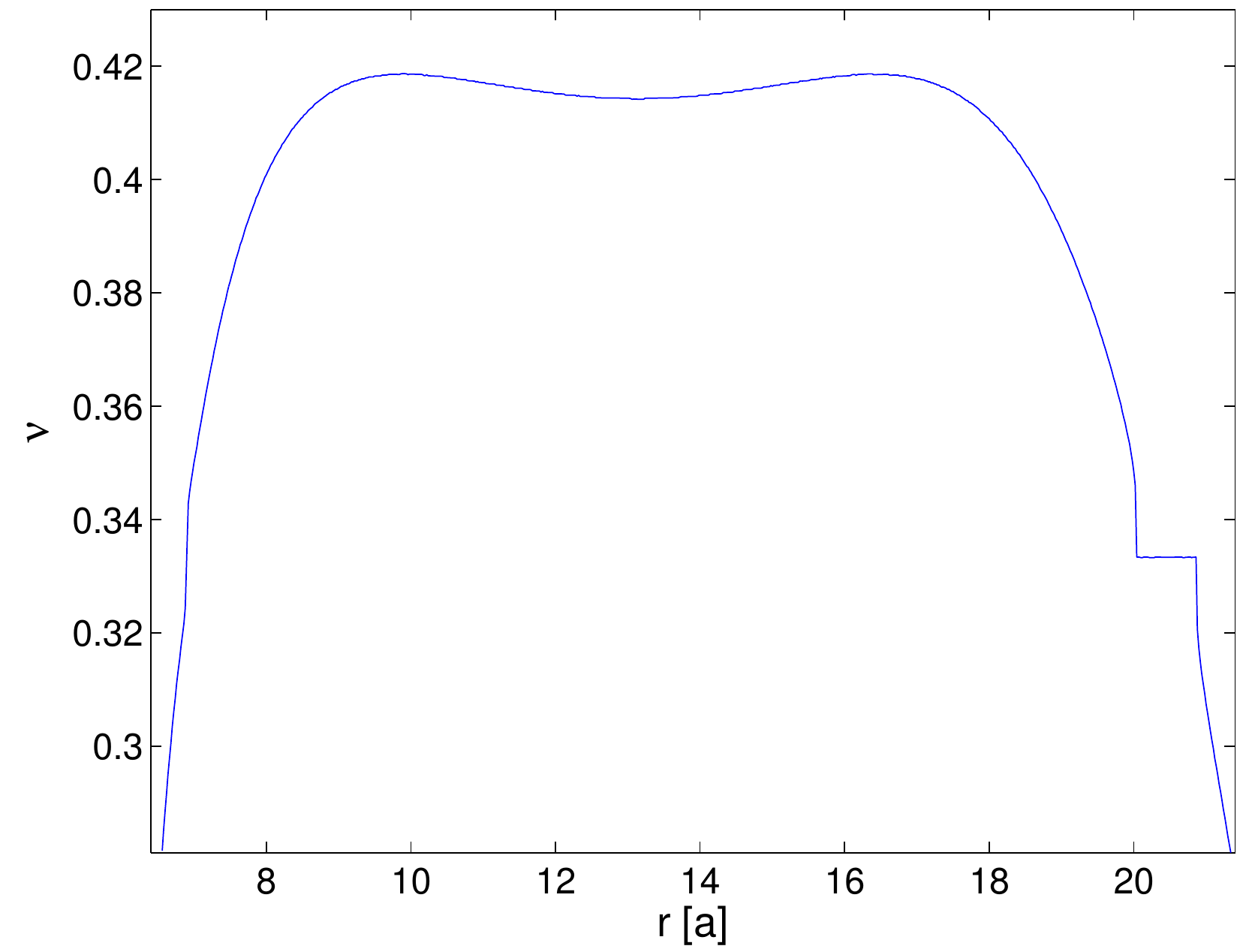}\\
\includegraphics[scale=0.44,trim=0mm 0mm 0mm 0mm,clip]{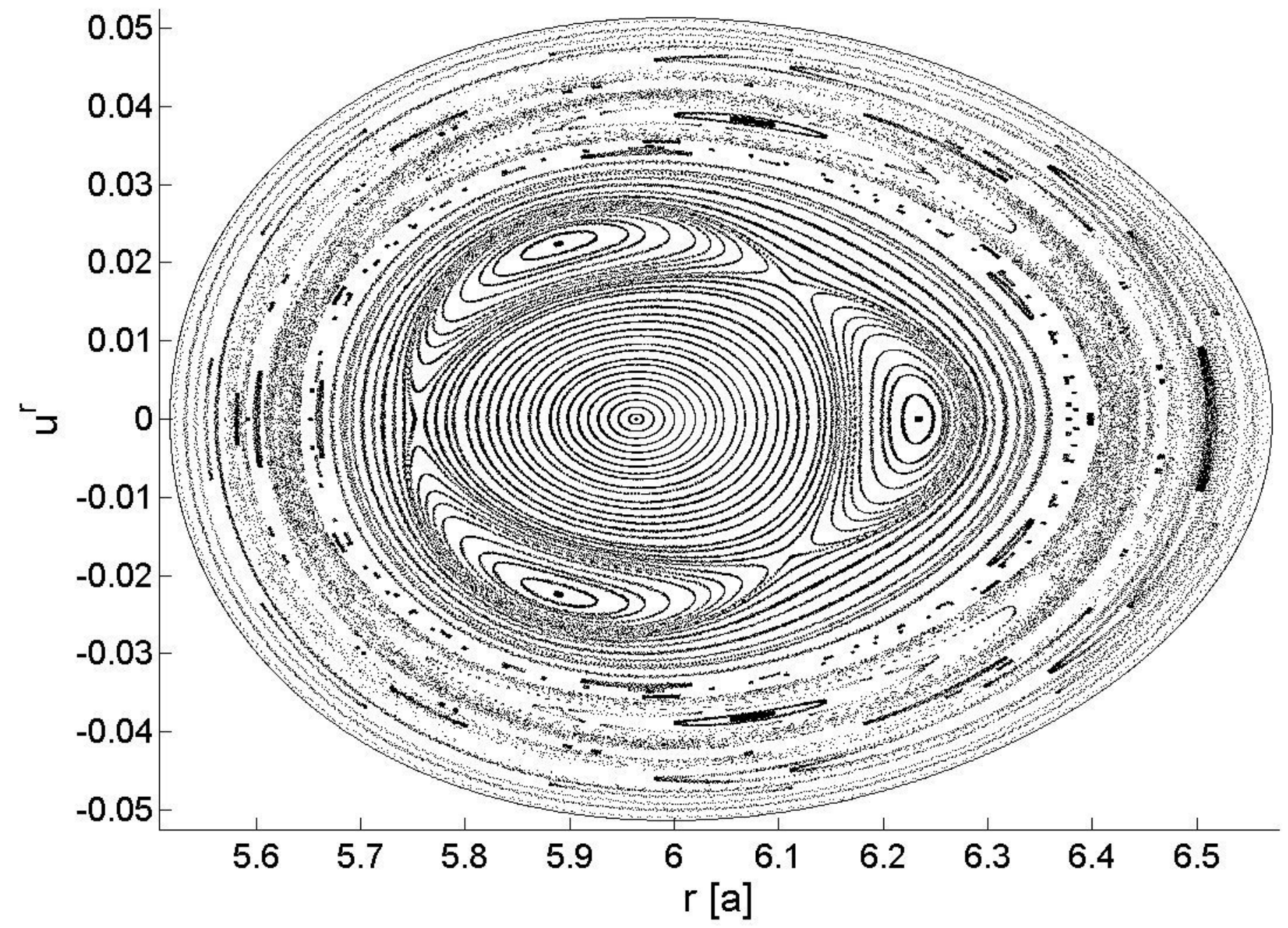}
\includegraphics[scale=0.41,trim=0mm 0mm 0mm 0mm,clip]{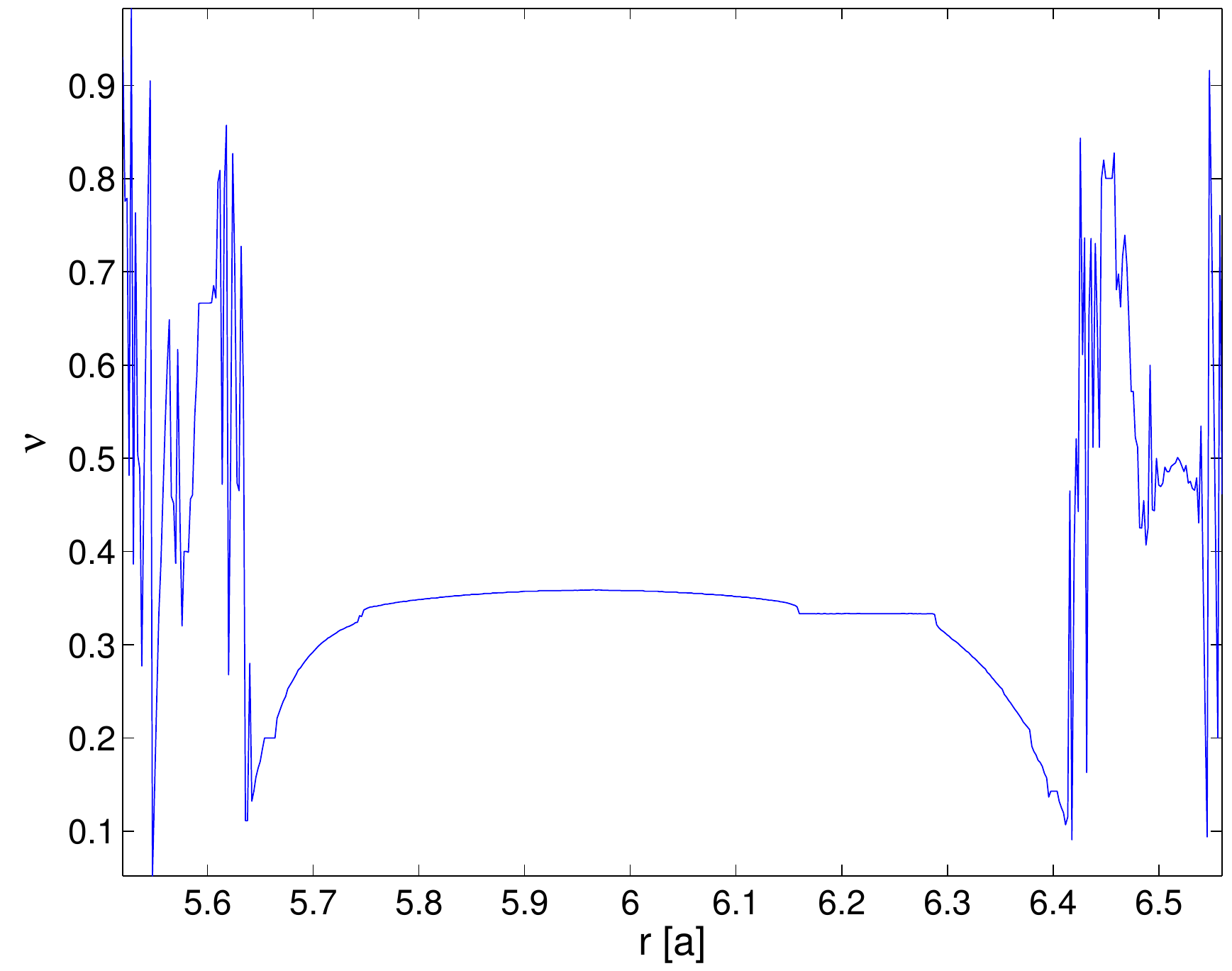}
\caption{Comparison of dynamics in equatorial lobes ($\theta_{\rm min}=\theta_{\rm sec}=\pi/2$). Left panels show equatorial Poincar\'e surfaces of section while the corresponding rotation numbers of trajectories are presented in the right ones. Common parameters of the trajectories in the top panels are $E=0.951$, $L=7.2058\,a$, $q=0$ and $b=0$ for which the equatorial potential minimum appears at $r_{\rm min}=15\,a$ with $V_{\rm min}=0.9494$. Middle panels show Poincar\'e section and rotation numbers for particles with $E=0.94$, $L=6.1076\,a$, $q=0$ and $b=2.8535\,a$ which brings the stable circular orbit to $r_{\rm min}=10\,a$ with $V_{\rm min}=0.9234$. Bottom panels are plotted for $E=0.818$, $L=-2.7277\,a$, $q=4.7181$ and $b=1\,a$  (while $V_{\rm min}=0.8165$ at $r_{\rm min}=6\,a$).}
\label{Fig:10}
\end{figure}

In order to compare dynamic properties of the system in all three cases (particle in non-magnetized $b=0$ spacetime, uncharged particle in magnetized spacetime and general case $q\neq0$, $b\neq0$), we first investigate the motion in equatorial potential wells, since there are no circular halo orbits for $b=0$. In figure \ref{Fig:10}, we present series of Poincar\'e surfaces of section with $\theta_{\rm sec}=\pi/2$ along with the corresponding plots of a rotation number as a function of initial value of radial coordinate. 
\begin{figure}[tbhp!]
\centering
\includegraphics[scale=0.44,trim=0mm 0mm 0mm 0mm,clip]{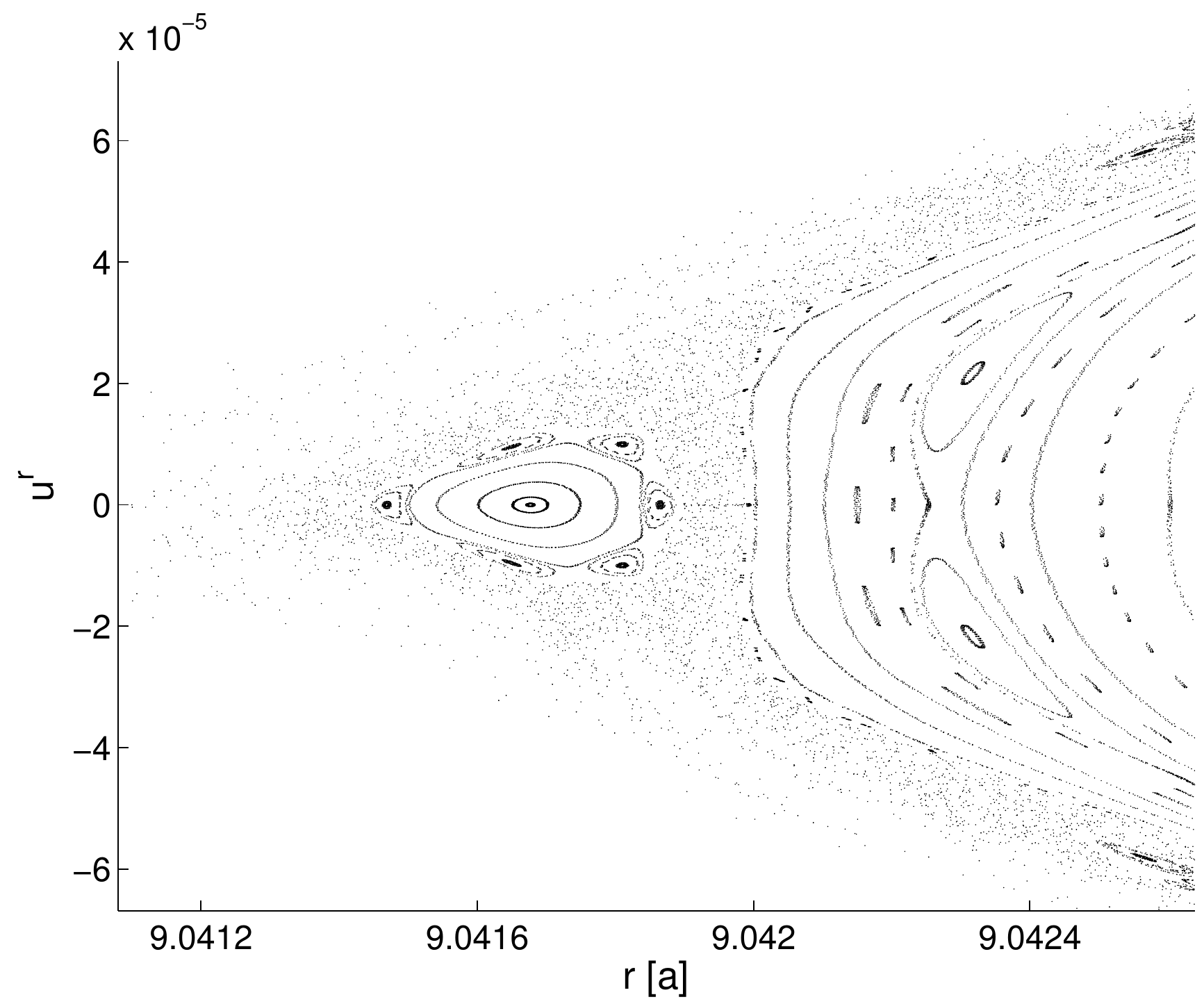}
\includegraphics[scale=0.43,trim=0mm 0mm 0mm 0mm,clip]{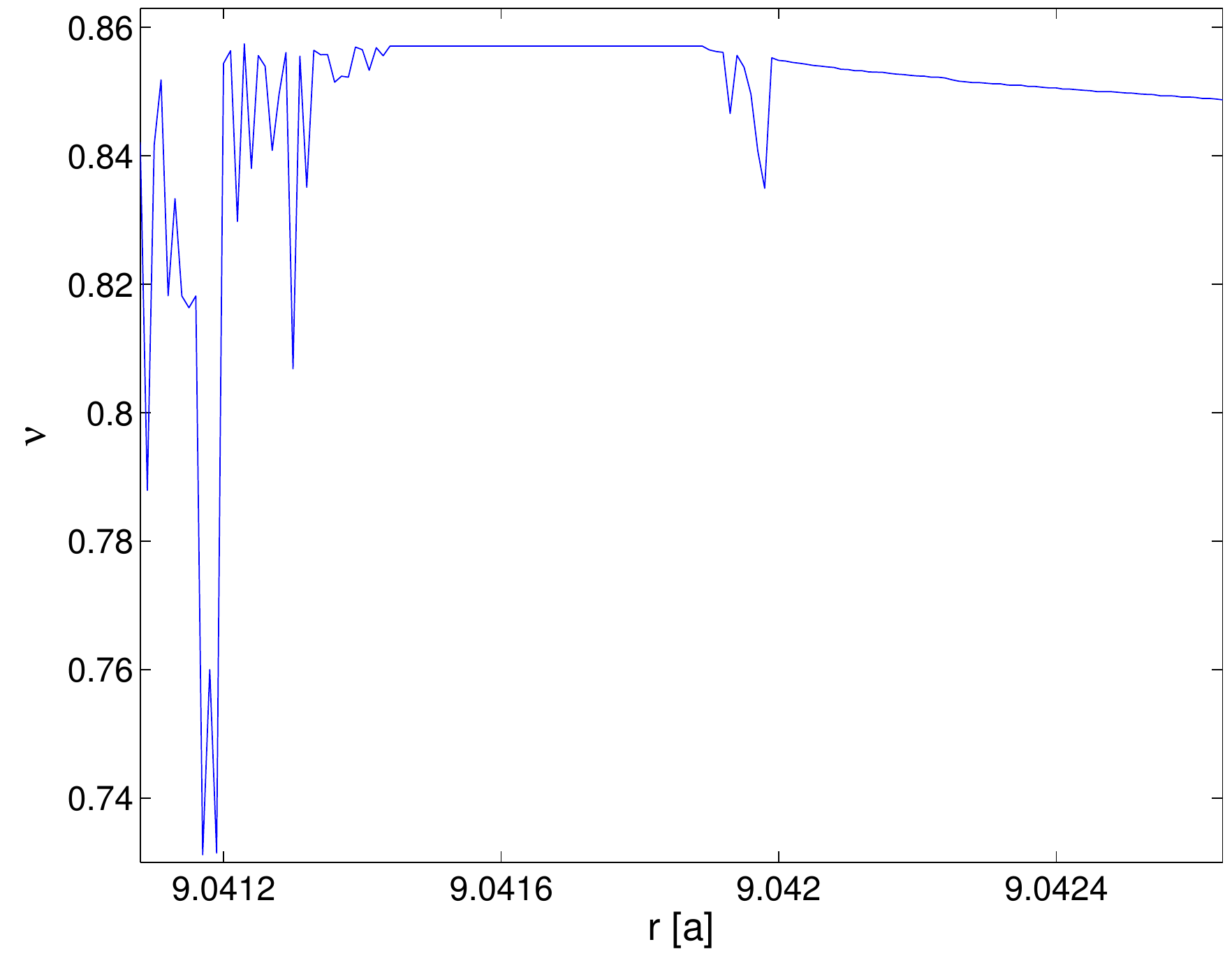}
\caption{Dynamics of test particles in non-magnetized Bonnor spacetime ($b=0$) near the throat of the equatorial lobe ($\theta_{\rm min}=\theta_{\rm sec}=\pi/2$). Left panel shows equatorial Poincar\'e surface of section, while the corresponding rotation number $\nu(r)$ of trajectories are presented in the right one. Common parameters of the trajectories in both the panels are $E=0.9522$, $L=7.2058\,a$, $q=0$ and $b=0$ for which the equatorial potential minimum appears at $r_{\rm min}=15\,a$ with $V_{\rm min}=0.9494$. Unlike the case presented in the upper panels of figure \ref{Fig:10}, here the potential lobe is opened allowing the particles to fall onto the horizon. The island of stability observed in the surface of section corresponds to the $\nu=6/7$ resonance.}
\label{Fig:10b}
\end{figure}

Upper left panel shows the Poincar\'{e} surface of the test particles trajectories when the magnetic dipole is switched off by setting $b=0$. We observe perfectly ordered motion with no traces of secondary fixed points nested in Birkhoff islands nor the chaotic orbits. Such simple pattern on the section is characteristic for integrable systems. The integrability conjecture is also supported by the behaviour of rotation number, which is smooth and non-constant throughout the lobe. While we show roughly $50$ trajectories on the section, the $\nu$-plot is constructed from around $1000$ of them, which allows much more detailed inspection for the possible presence of tiny chaotic domains or faint resonances. 

If we perturb the system by the magnetic field, however, the chain of Birkhoff islands develops. Although we know that there are some integrable system with resonant islands of single multiplicity \cite{contopoulos02}, here its presence arouses suspicion of nonintegrability, since none were present for $b=0$. Indeed, in the following (see figure \ref{Fig:11}), we observe chaotic motion in this setup ($q=0$, $b\neq0$), which is irrefutable evidence of being nonintegrable. 

Finally, in the bottom panel of figure \ref{Fig:10}, we introduce the charge of the test particle. We choose such a combination of parameters that leads to the same value of ratio $r_{\rm min}/r_{\rm h}$, as it is acquired in the previous uncharged case. This makes the two cases better comparable and the effect of the newly introduced electromagnetic forces more distinguishable. 
In the last surface of section, we really observe much more complex patterns compared to those of uncharged particles. KAM curves of quasiperiodic orbits are present as well as several Birkhoff chains of islands corresponding to the resonances of intrinsic frequencies of the system. These are interwoven with pronounced chaotic layers. Such a picture is typical for considerably perturbed system far from integrability. 

The presented sequence of Poincar\'{e} surfaces of section suggests the following: the motion of the test particles in the Bonnor spacetime without magnetic dipole ($b=0$) appears perfectly regular and therefore it might be integrable, while the magnetic parameter $b$ acts as a nonintegrable perturbation here (indicated by the chaos observed in figure \ref{Fig:11}). With the introduction of ionised test particles, however, the system departs considerably farther away from the integrability.
 
The conjecture of integrability of test particle motion in the vicinity of non-magnetized Bonnor's solution (which can actually be identified with $\delta=2$ member of the class of exact Zipoy-Voorhees solutions \cite{zipoy66,voorhees70}) deserves further examination. We first check throughout the parameter space to verify that the dynamics in the closed potential lobes always follows the simple regular pattern observed in the upper panels of  figure \ref{Fig:10}. Detailed numerical study revealed no traces of chaos and would therefore support the claim of integrability. Besides these numerical indications, we also recall that Bonnor obtained the static magnetic dipole solution \cite{Bonnor66} by transforming the Kerr stationary metric in which the test particle motion is known to be integrable \cite{carter68}. 

However, if we turn our attention from the closed potential lobes to the vicinity of saddle points (around which the throat allowing the particle to escape the well and fall onto the horizon develops), we eventually observe tiny chaotic layers as illustrated in figure \ref{Fig:10b}. These chaotic orbits correspond to those particles which actually leave the potential well after certain amount of time. Besides the small fraction of escaping particles which exhibit chaotic behavior, the majority of orbits in the lobe remains stable and regular. Nevertheless, the presence of chaotic orbits represents a conclusive evidence of nonintegrability of a given system. Coincidentally, the nonintegrability of motion in a general Zipoy-Voorhees spacetime was very recently shown by Lukes-Gerakopoulos \cite{lukes12}, where the issue was treated in detail. Therefore, we end the discussion here and summarize that although the dynamics in the non-magnetized Bonnor spacetime (i.e. Zipoy-Voorhees with $\delta=2$) is typically regular (we actually found no chaotic orbit in closed lobes), the underlying system is not integrable.

\section{Motion in halo lobes}
\begin{figure}[ht]
\centering
\includegraphics[scale=0.42,trim=0mm 0mm 0mm 0mm,clip]{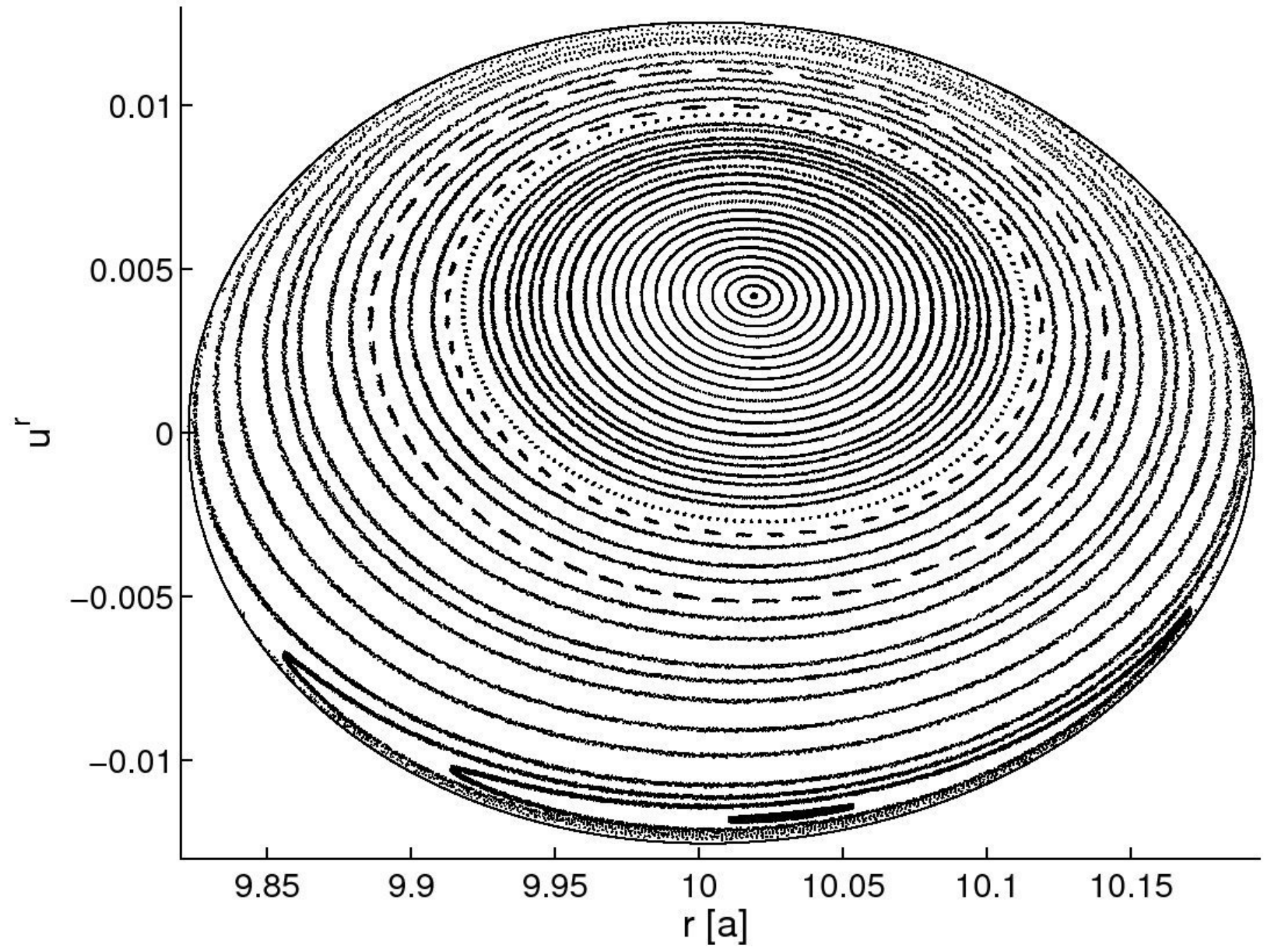}
\includegraphics[scale=0.42,trim=0mm 0mm 0mm 0mm,clip]{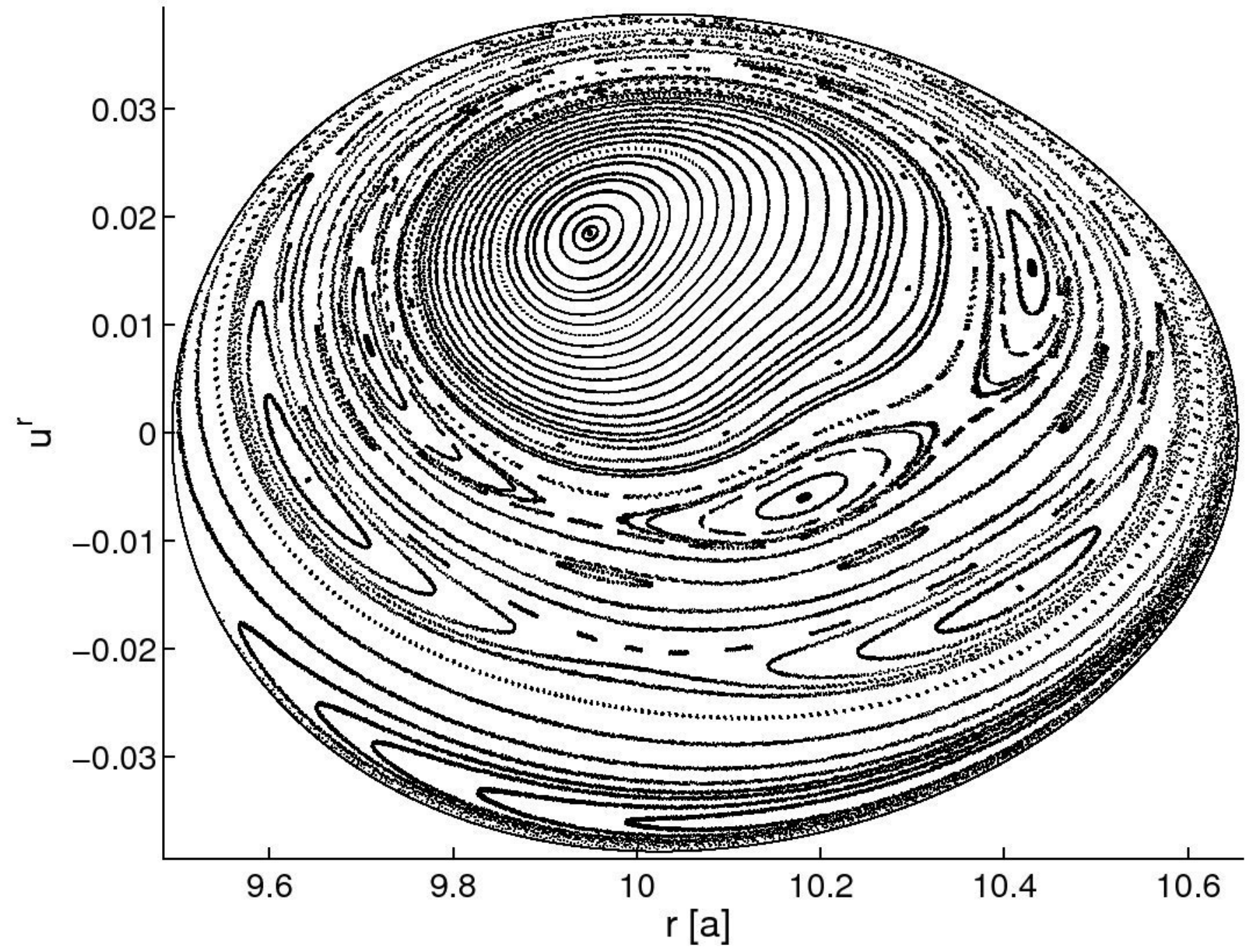}
\includegraphics[scale=0.42,trim=0mm 0mm 0mm 0mm,clip]{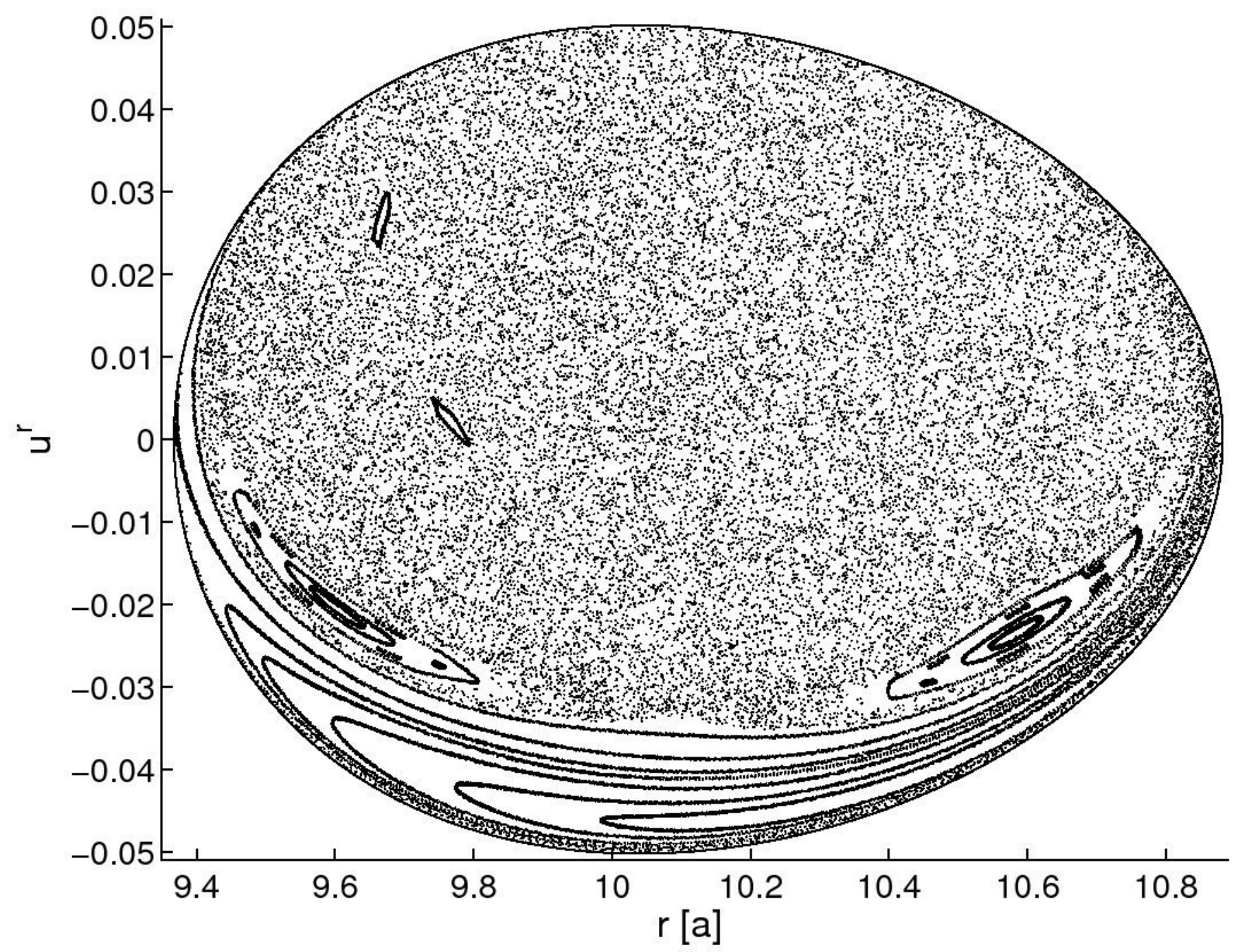}
\includegraphics[scale=0.42,trim=0mm 0mm 0mm 0mm,clip]{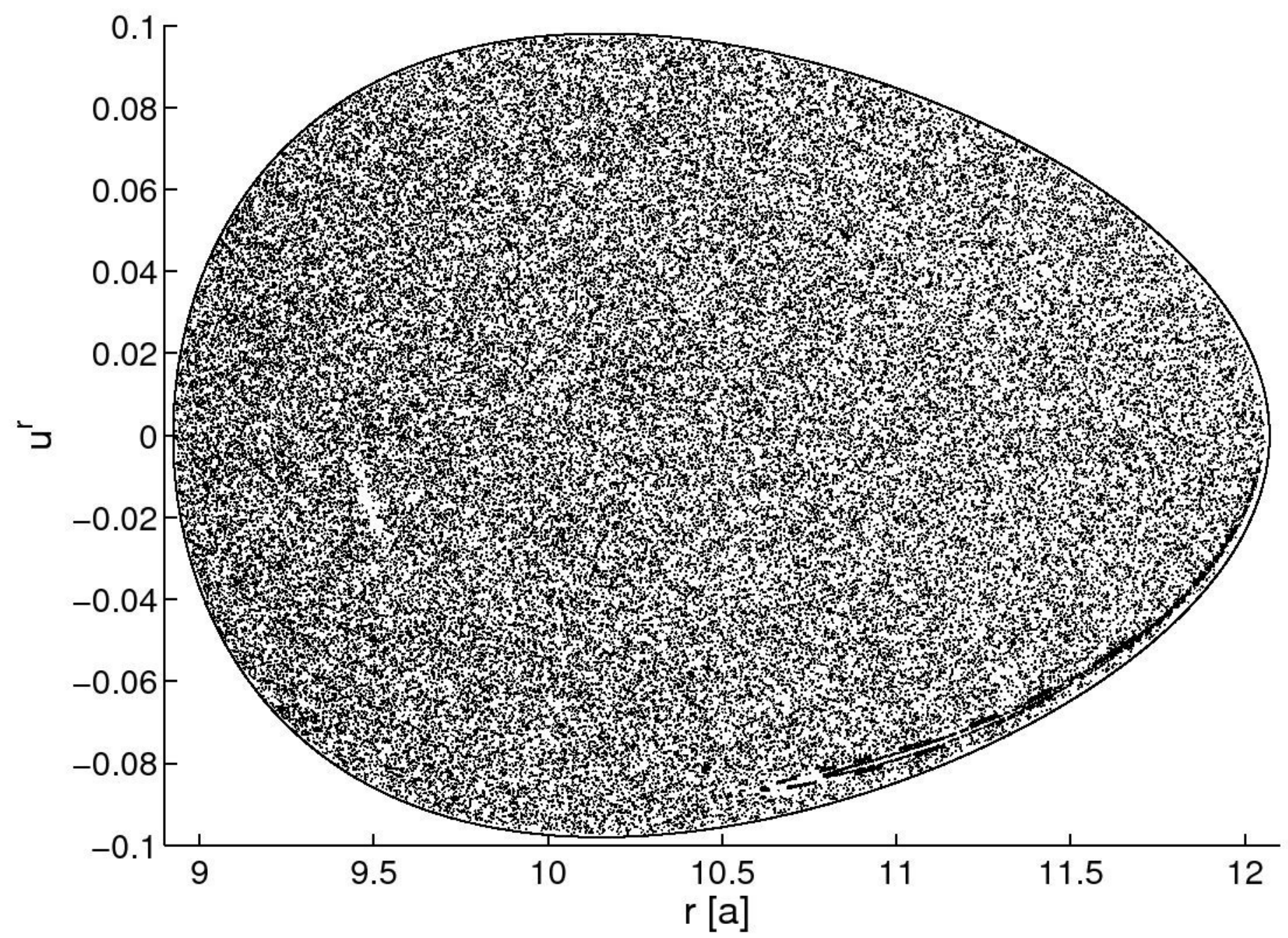}
\caption{Poincar\'{e} surfaces of section ($\theta_{\rm sec}=\pi/3$) for uncharged particles moving within halo lobes in the Bonnor spacetime with $b=5.9771\,a$. Particles with $L=3.6743\,a$ are launched from a vicinity of the off-equatorial potential well minimum ($r_{\rm min}=10\,a$, $\theta_{\rm min}=\theta_{\rm sec}=\pi/3$ and $V_{\rm min}=0.8717$) with various values of energy. The upper left panel shows the section for the level $E=0.8718$ (small halo lobe), in the upper right we set $E=0.873$ (large halo lobe), $E=0.8739$ produces cross-equatorial lobe which just emerged from symmetric halo lobes (bottom left panel), while with $E=0.88$ we obtain the large cross-equatorial lobe which almost opens.}
\label{Fig:11}
\end{figure}

\begin{figure}[ht]
\centering
\includegraphics[scale=0.412,trim=0mm 0mm 0mm 0mm,clip]{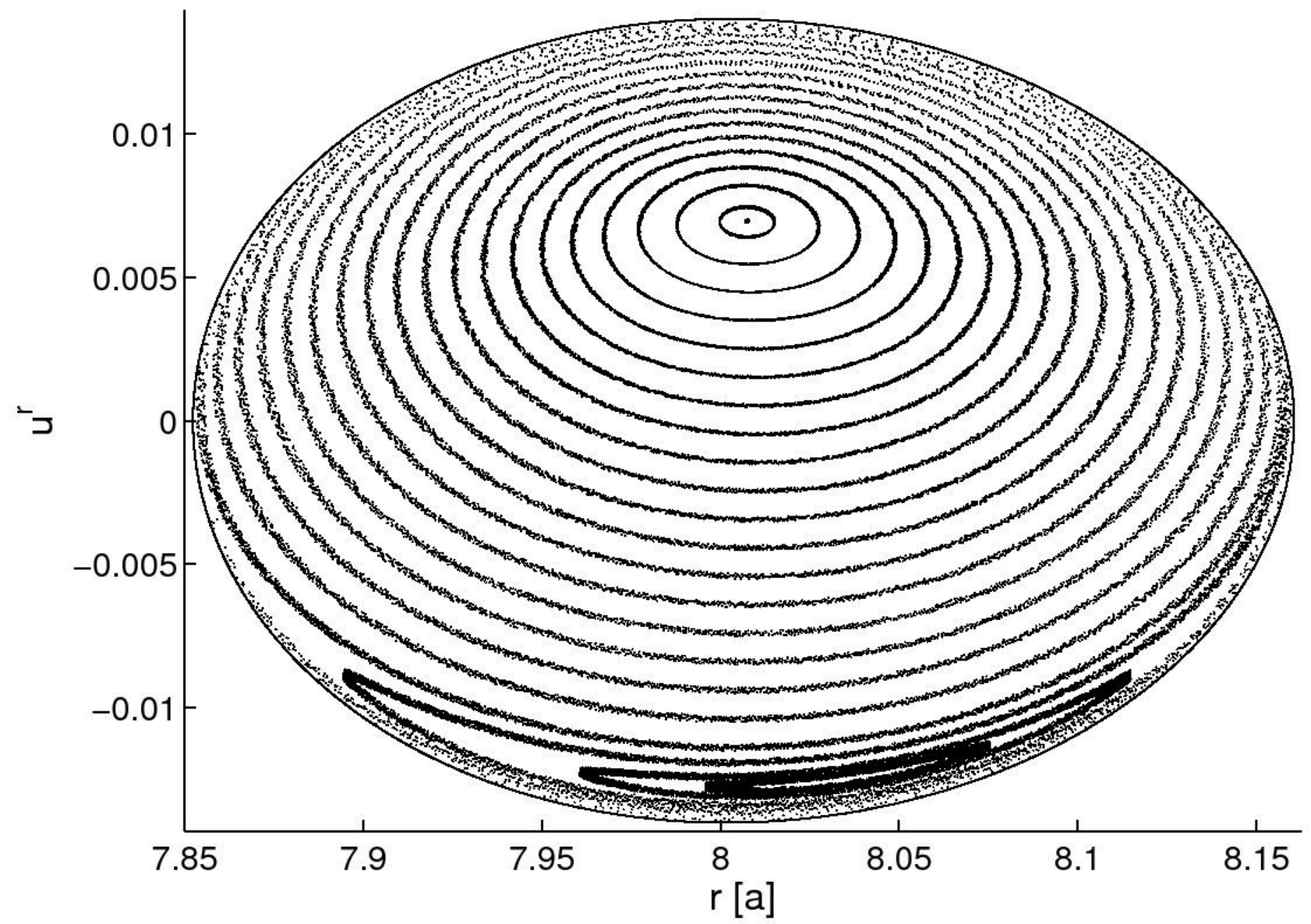}
\includegraphics[scale=0.412,trim=0mm 0mm 0mm 0mm,clip]{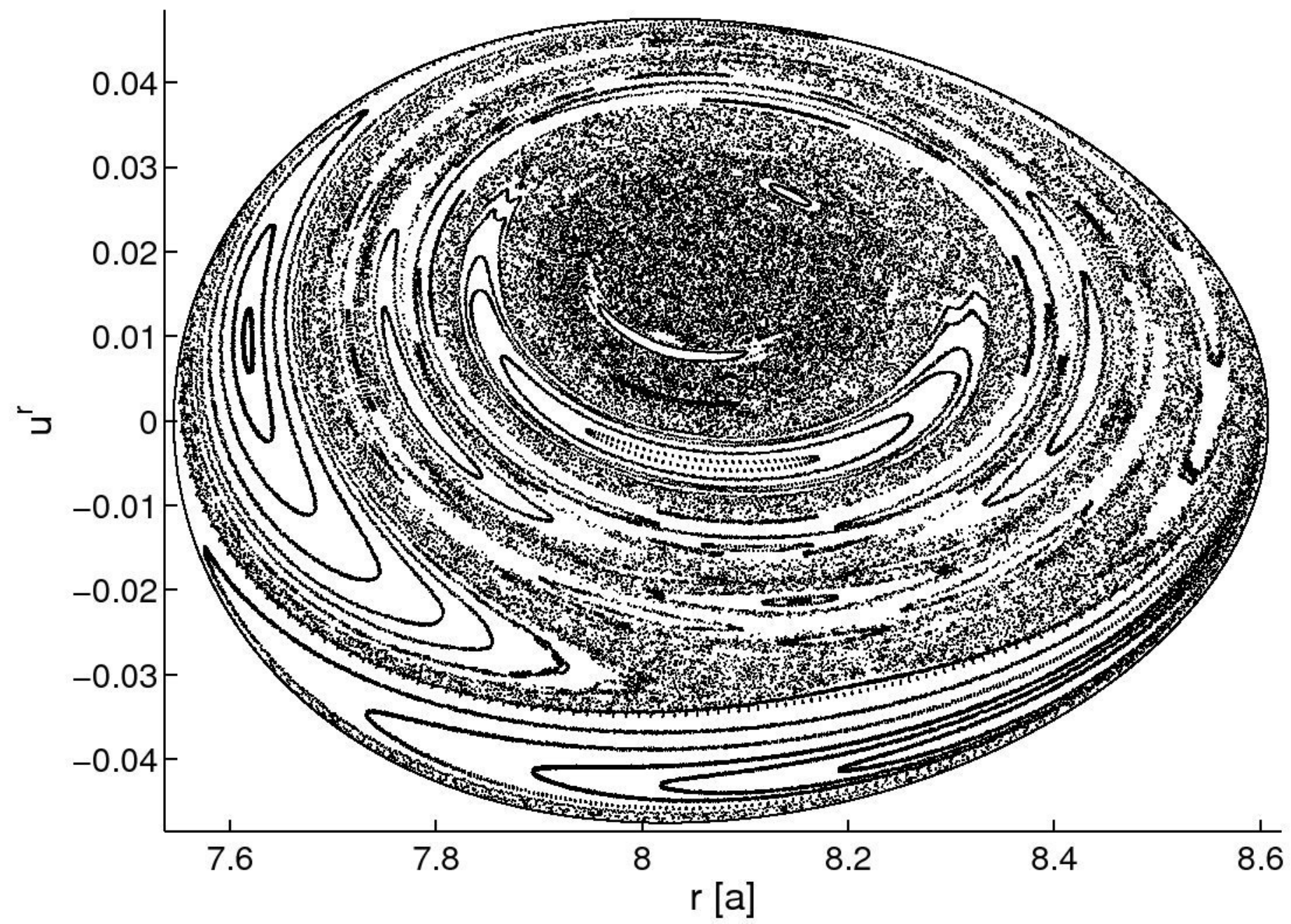}
\includegraphics[scale=0.412,trim=0mm 0mm 0mm 0mm,clip]{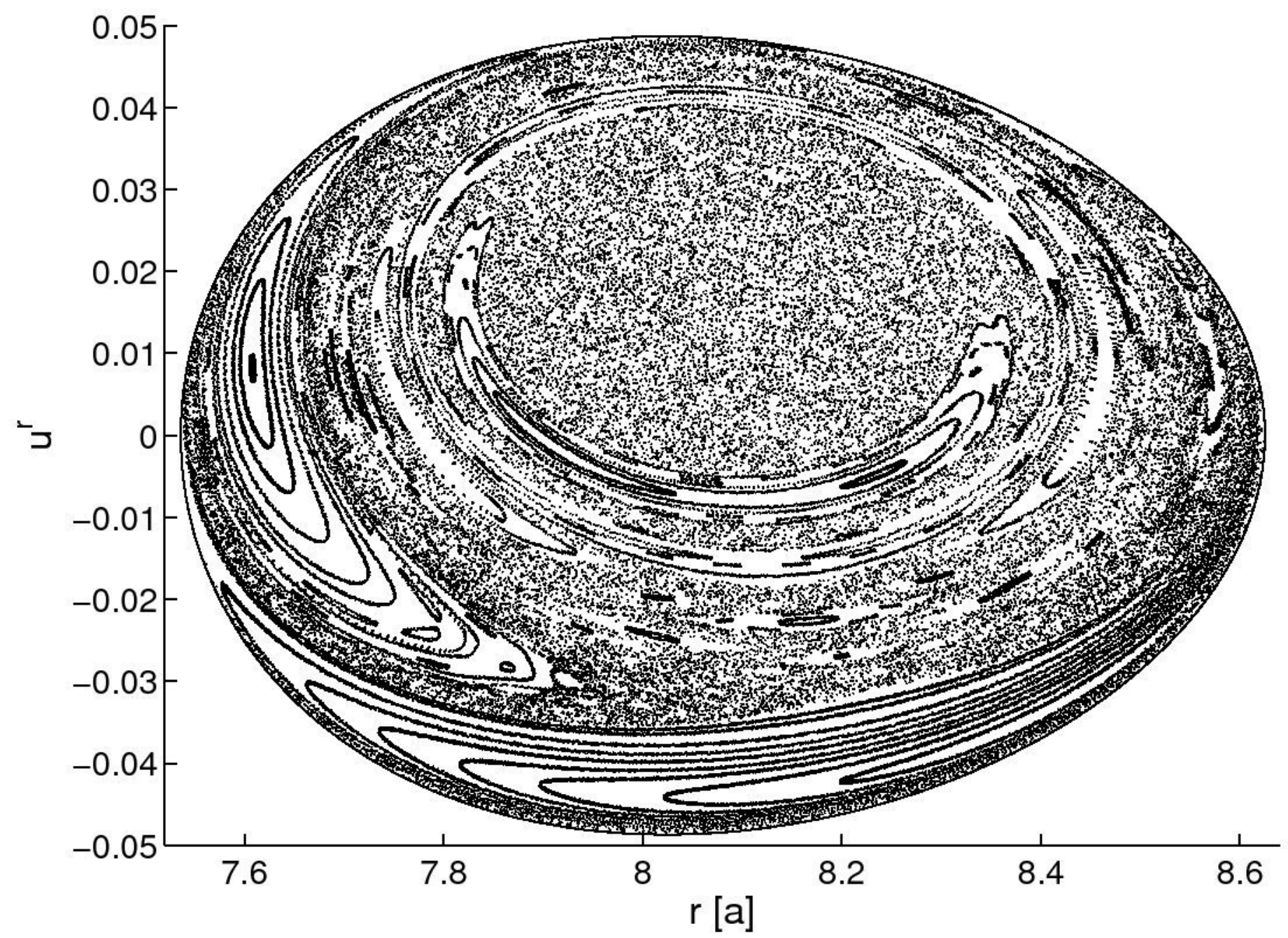}
\includegraphics[scale=0.412,trim=0mm 0mm 0mm 0mm,clip]{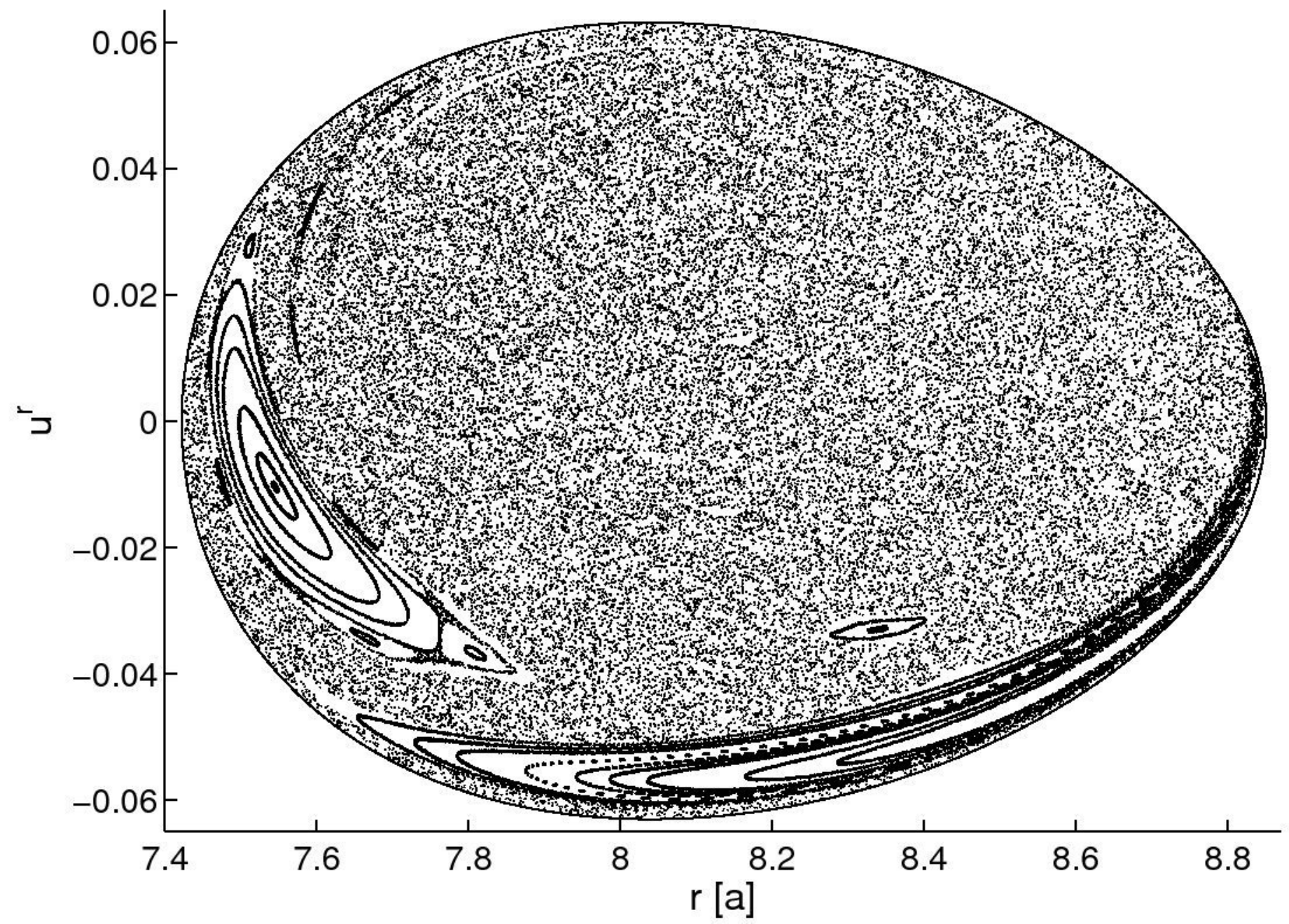}
\caption{Poincar\'{e} surfaces of section ($\theta_{\rm sec}=\pi/3$) for charged particles with $q=0.1259$ moving within halo lobes in the Bonnor spacetime with $b=4.5393\,a$. Particles with $L=-3.5486\,a$ are launched from a vicinity of the off-equatorial potential well minimum ($r_{\rm min}=8\,a$, $\theta_{\rm min}=\theta_{\rm sec}=\pi/3$ and $V_{\rm min}=0.8475$) with various values of energy. The upper left panel shows the section for the level $E=0.8477$ (small halo lobe), in the upper right we set $E=0.8495$ (large halo lobe), $E=0.8496$ produces cross-equatorial lobe which just emerged from symmetric halo lobes (bottom left panel) while with $E=0.851$ we obtain the large cross-equatorial lobe.}
\label{Fig:12}
\end{figure}

In the following, we compare dynamics in off-equatorial potential wells for uncharged (figure \ref{Fig:11}) and charged particles (figure \ref{Fig:12}). For the sake of better comparability, these are both chosen to have $\theta_{\rm min}=\theta_{\rm sec}=\pi/3$ and equal value of $r_{\rm min}/r_{\rm h}$. Both series show Poincar\'{e} surfaces of section of particles being launched from the vicinity of off-equatorial potential minima (in which the circular halo orbit resides) differing in energy $E$, which governs the size and shape of the lobe ($E$ sets the level at which the effective potential surface is being intersected).
\begin{figure}[hp!]
\begin{center}
\includegraphics[scale=0.28,trim=0mm 0mm 0mm 0mm,clip]{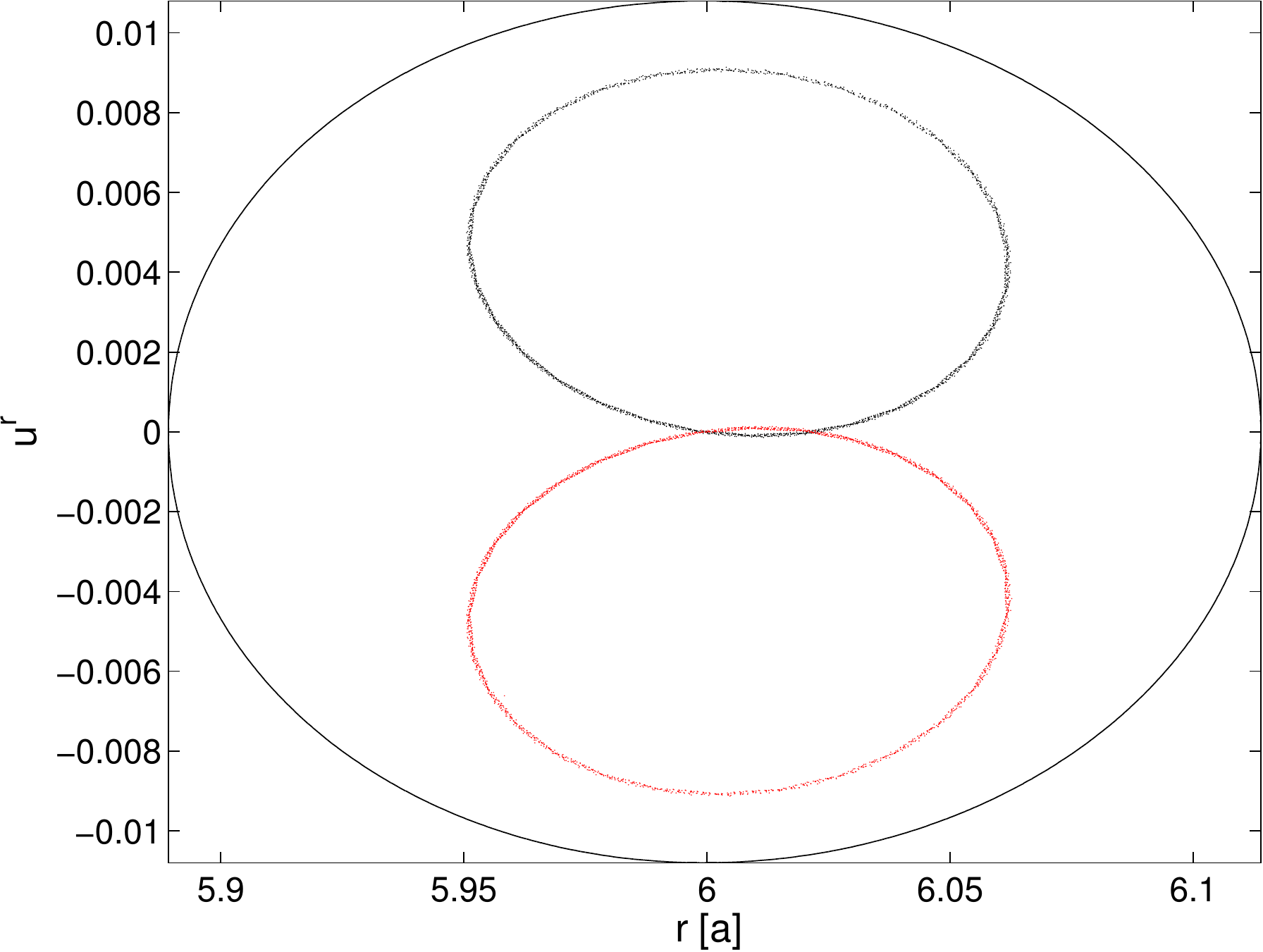}
\includegraphics[scale=0.28,trim=0mm 0mm 0mm 0mm,clip]{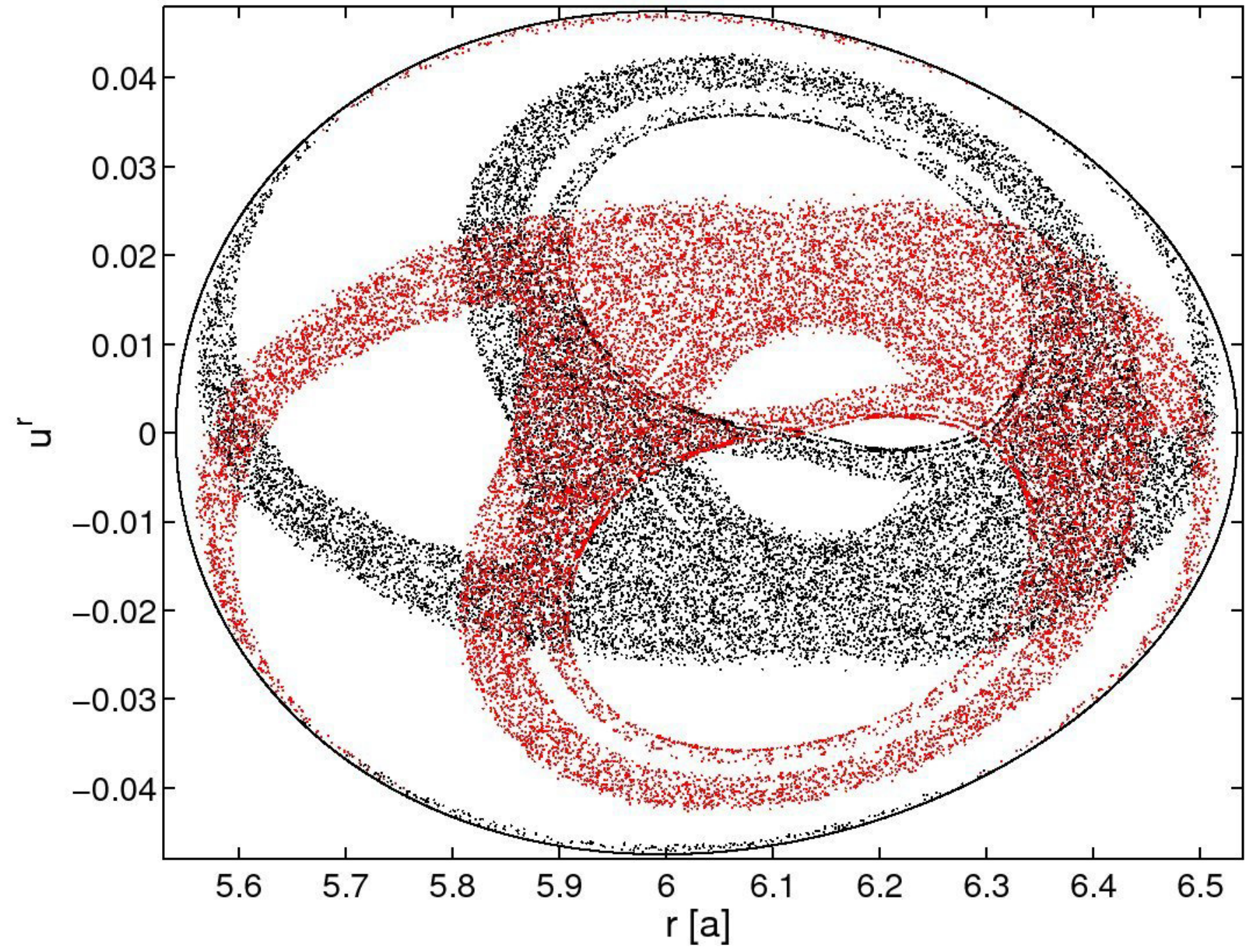}
\includegraphics[scale=0.28,trim=0mm 0mm 0mm 0mm,clip]{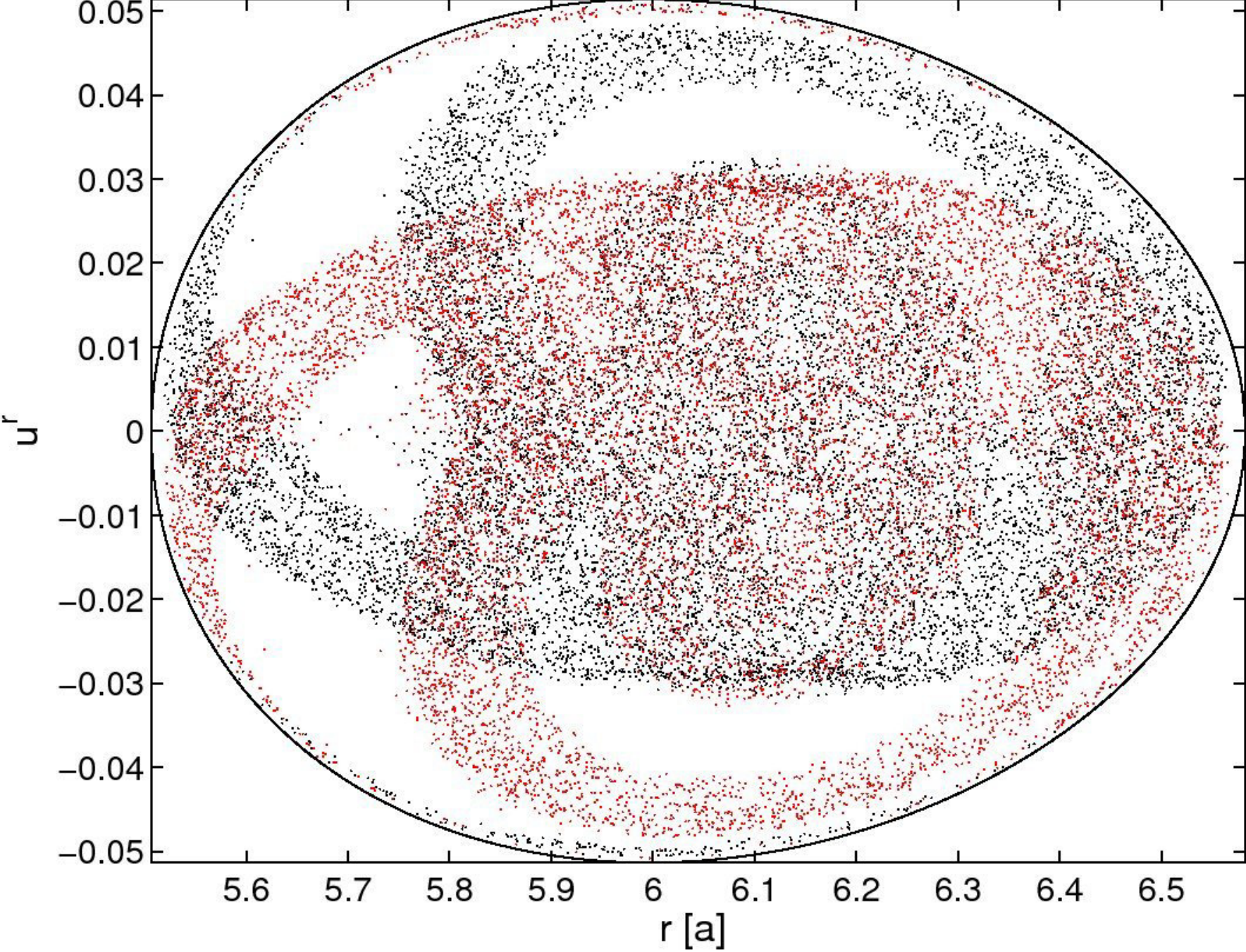}\\
\includegraphics[scale=0.31,trim=0mm 0mm 0mm 0mm,clip]{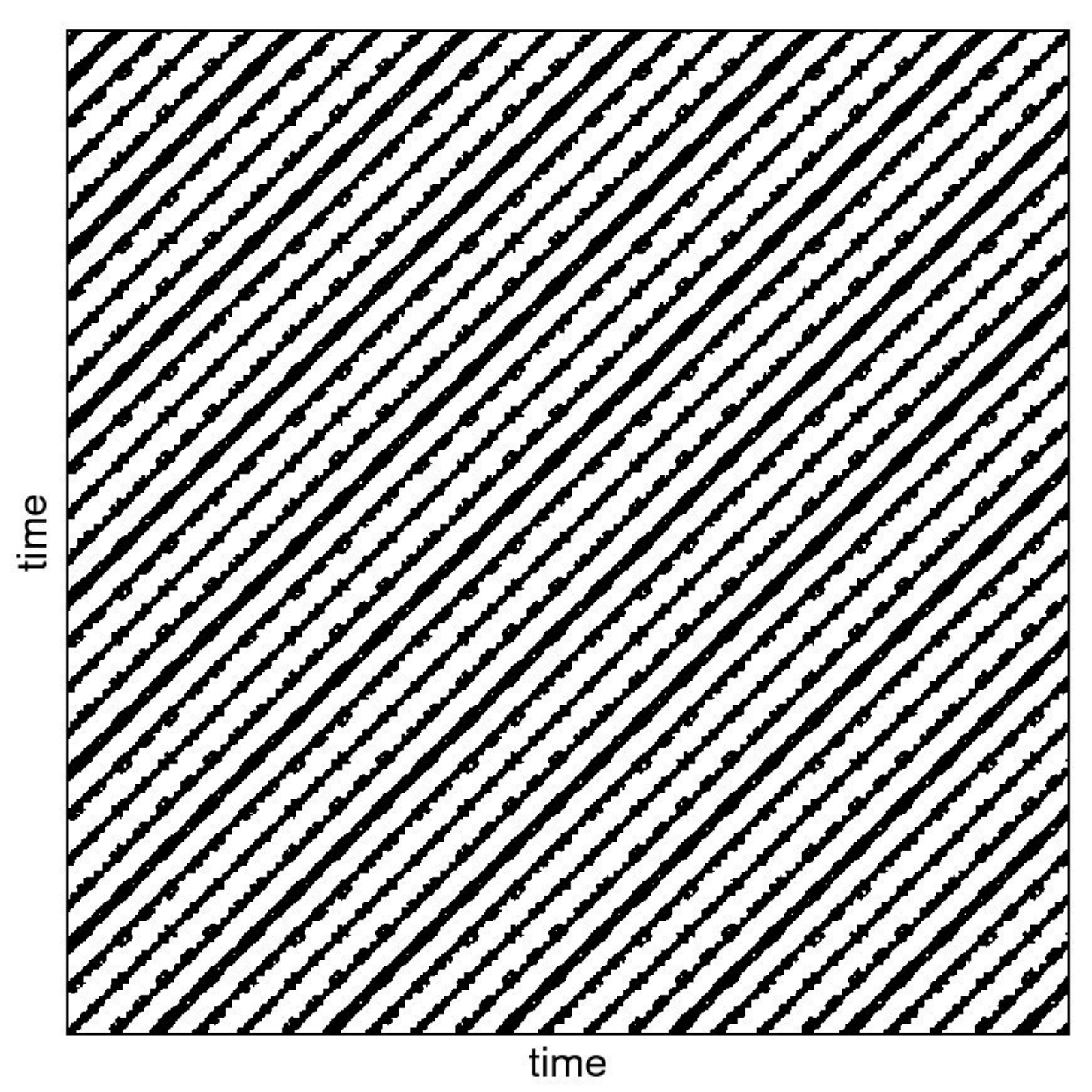}\hspace*{7mm}
\includegraphics[scale=0.32,trim=0mm 0mm 0mm 0mm,clip]{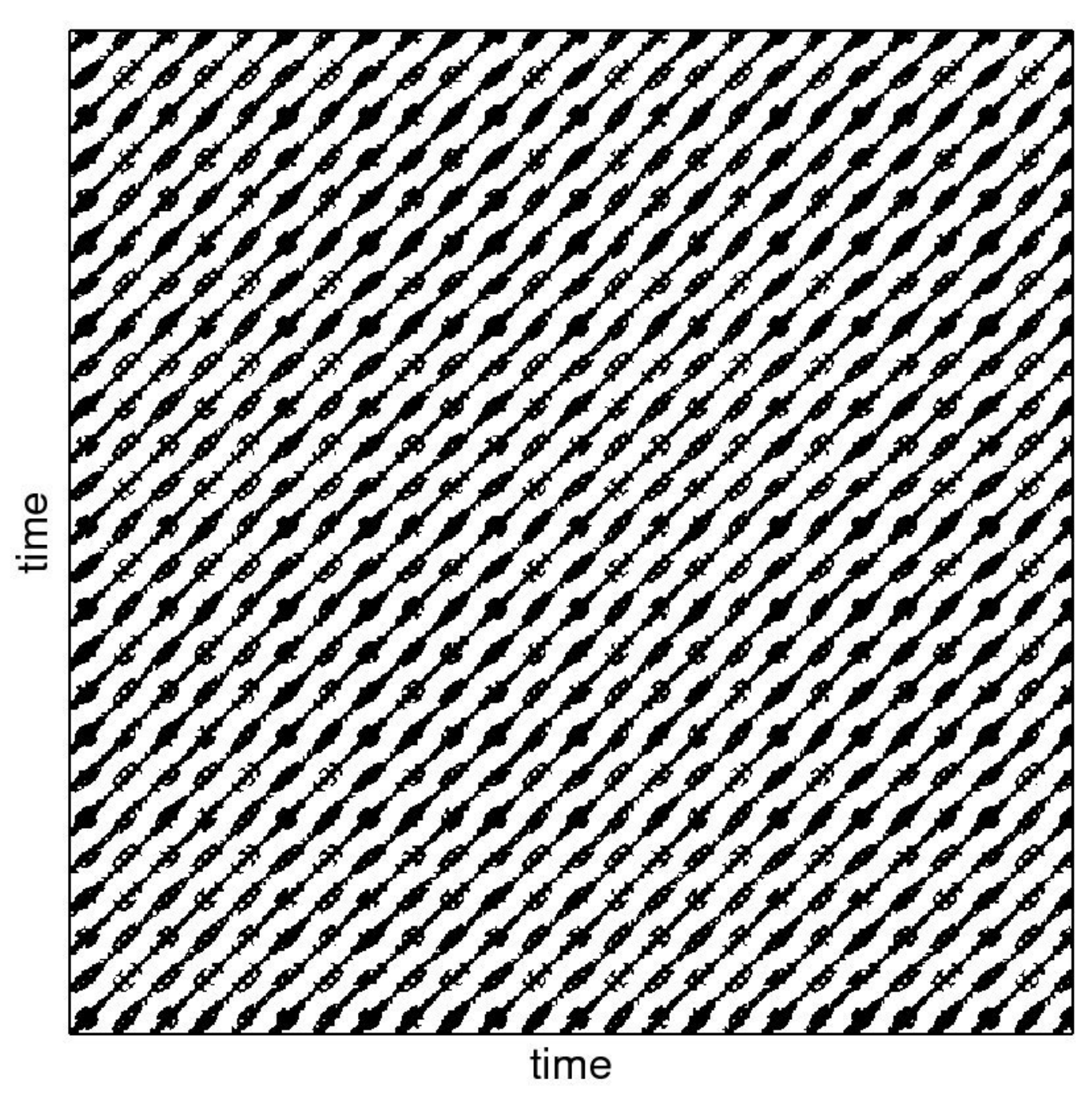}\hspace*{5mm}
\includegraphics[scale=0.295,trim=0mm 0mm 0mm 0mm,clip]{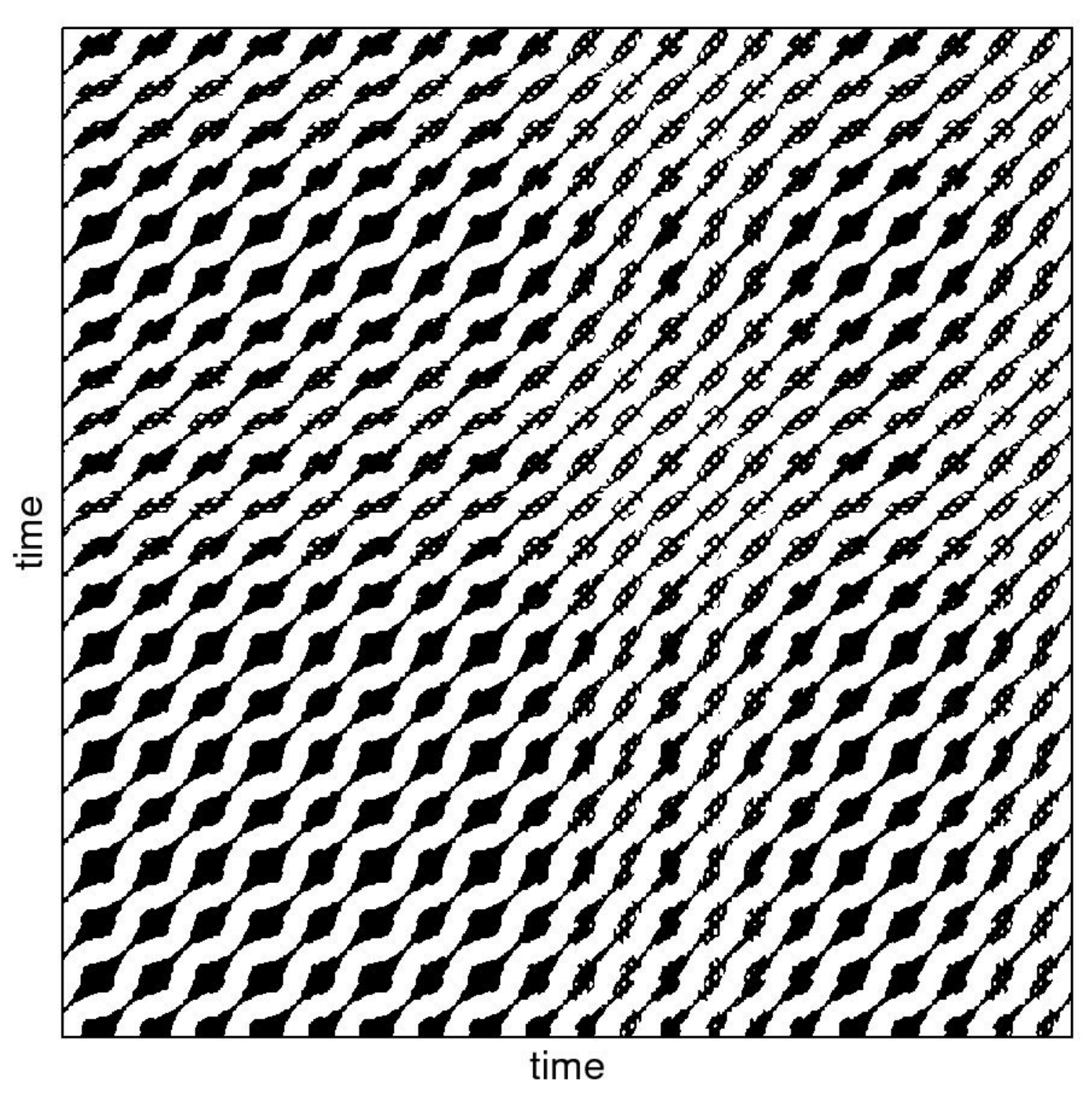}\\
\end{center}
\includegraphics[scale=0.28,trim=0mm 0mm 0mm 0mm,clip]{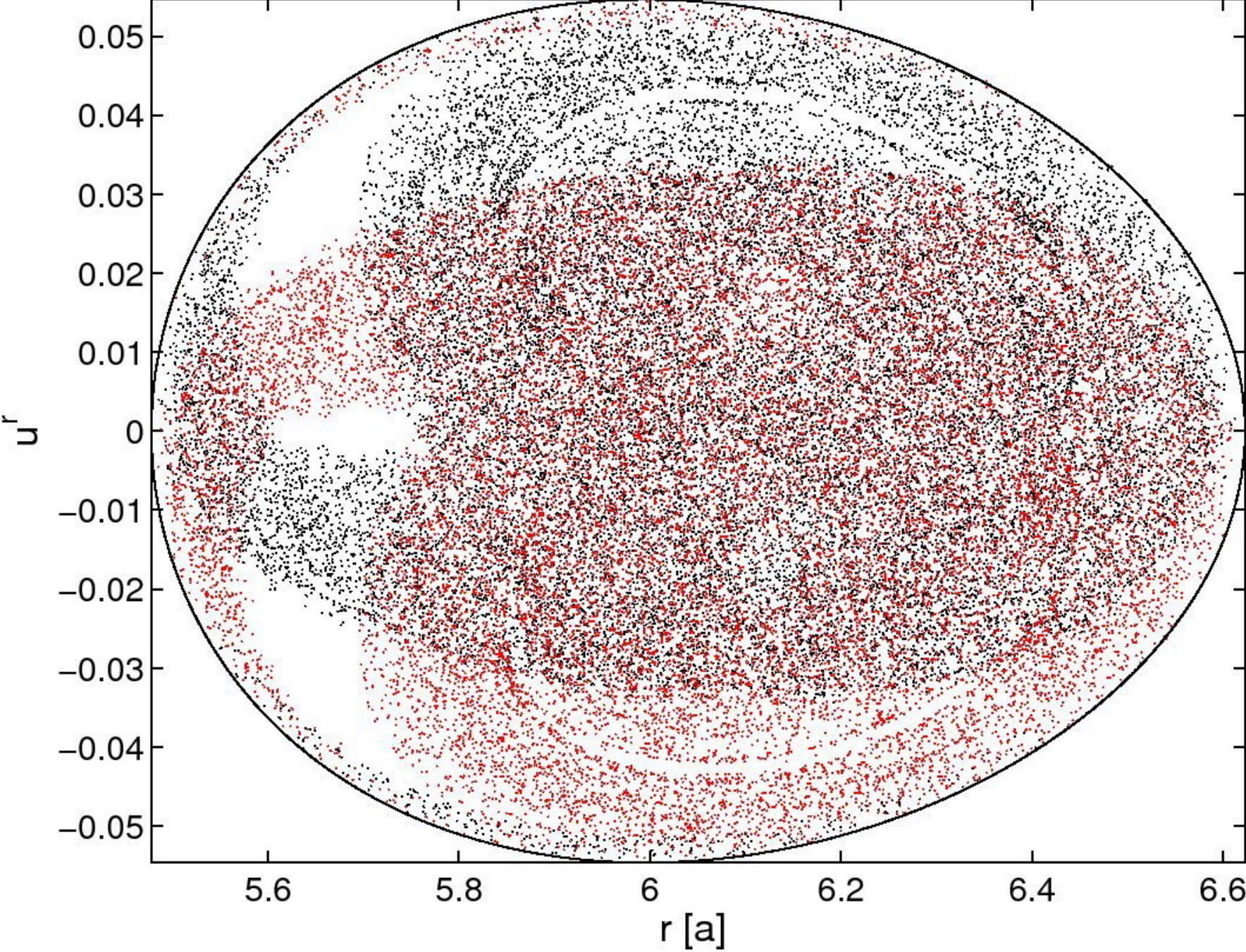}~
\includegraphics[scale=0.28,trim=0mm 0mm 0mm 0mm,clip]{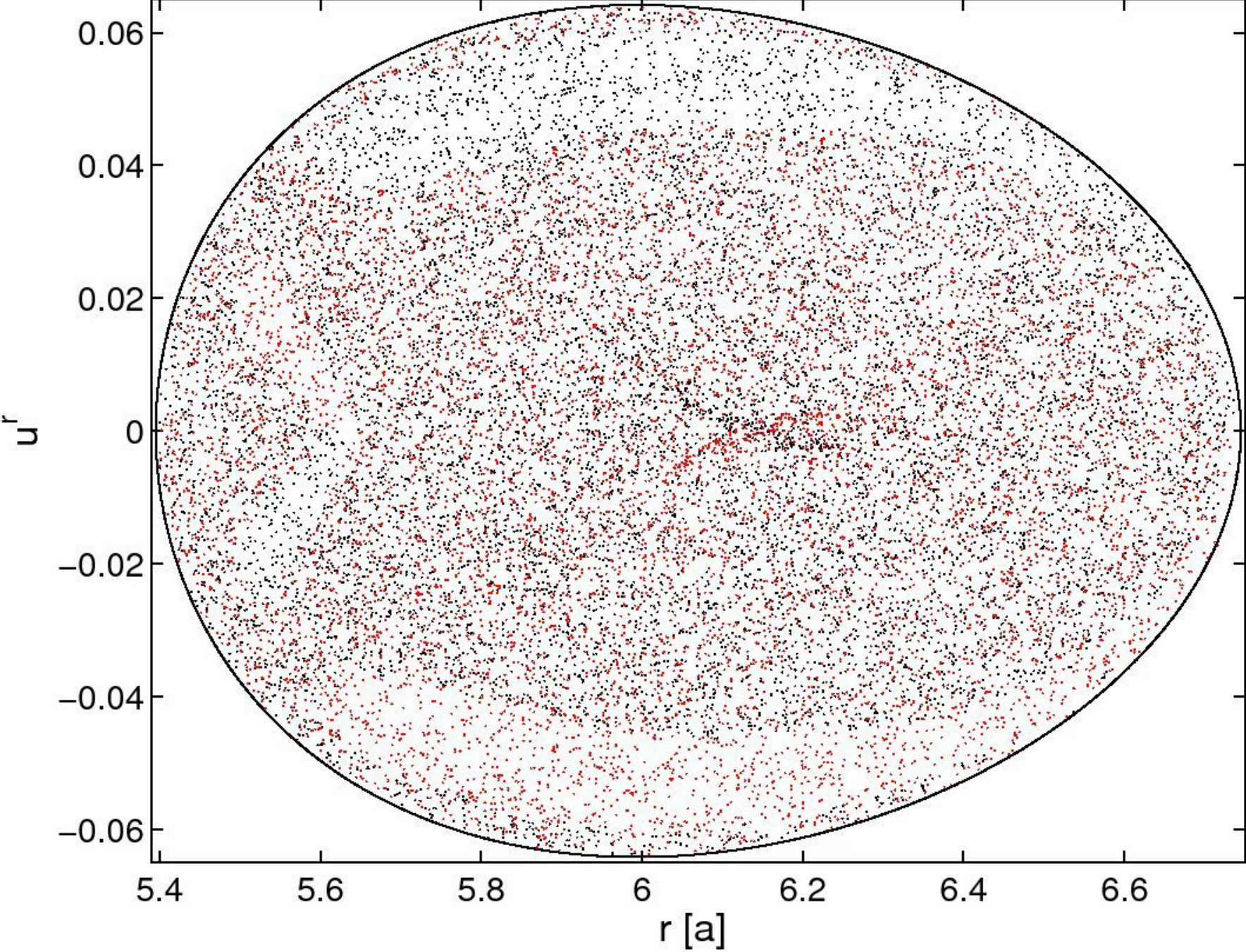}
\includegraphics[scale=0.28,trim=0mm 0mm 0mm 0mm,clip]{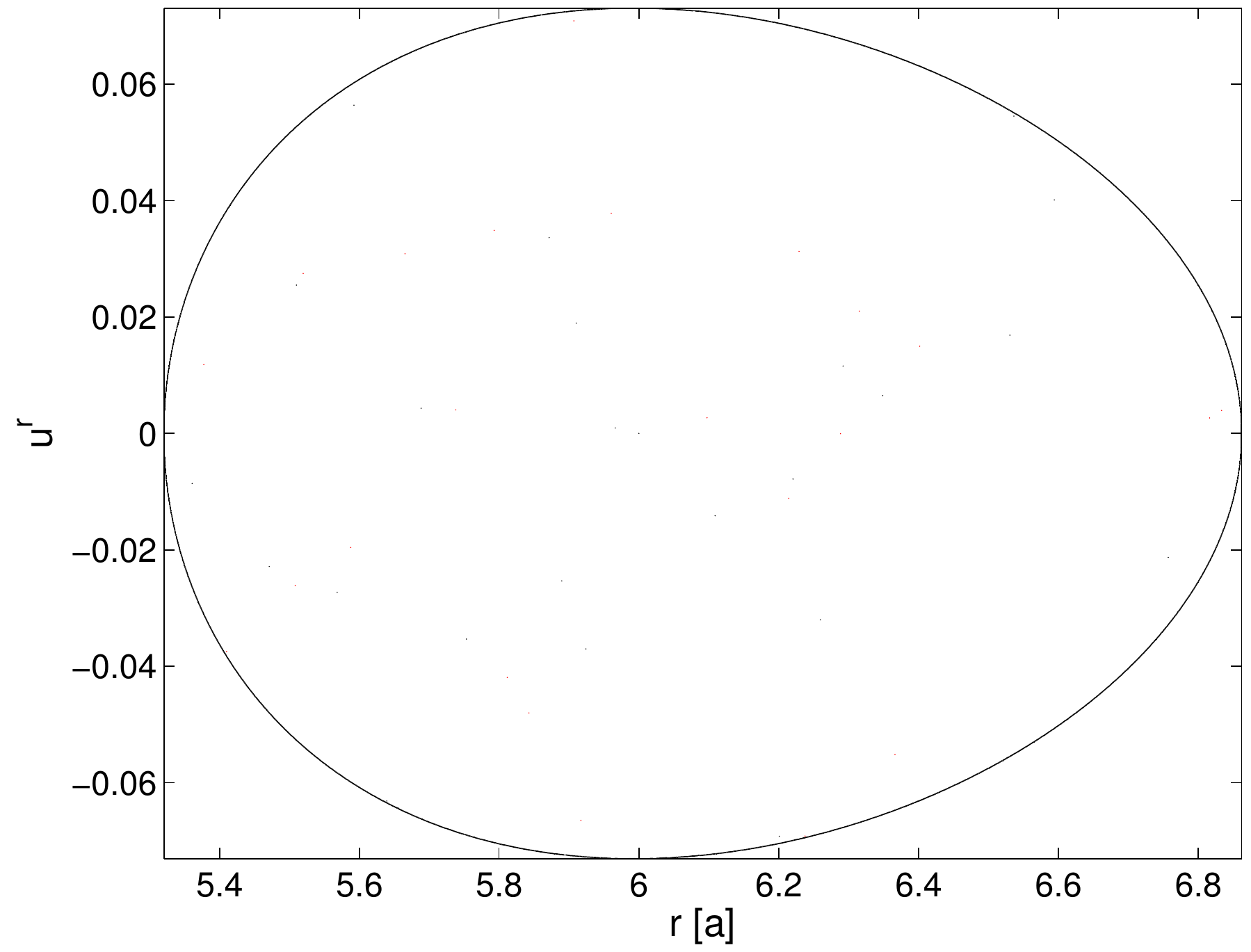}\\
\centering
\includegraphics[scale=0.3,trim=0mm 0mm 0mm 0mm,clip]{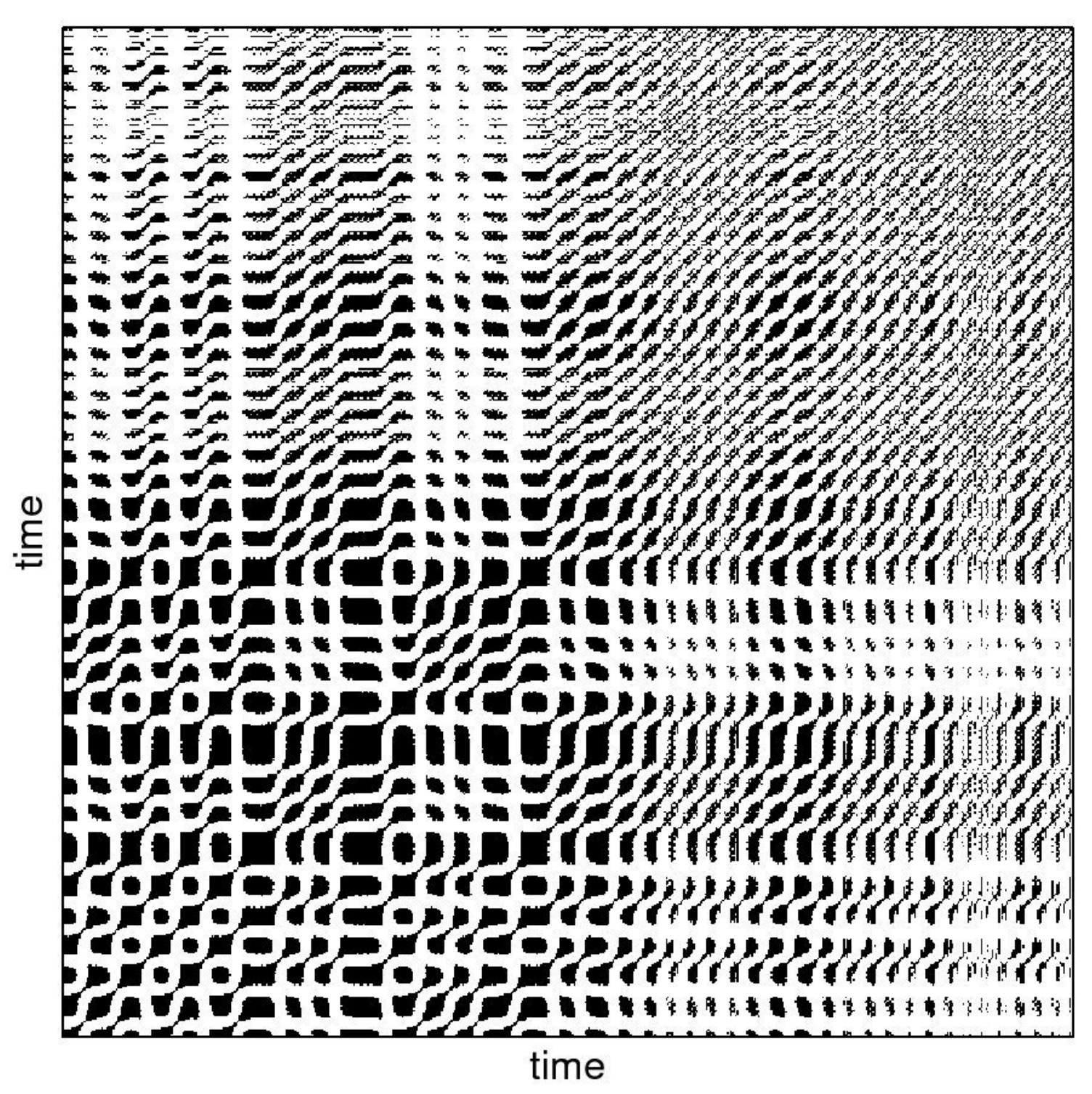}~~~~
\includegraphics[scale=0.31,trim=0mm 0mm 0mm 0mm,clip]{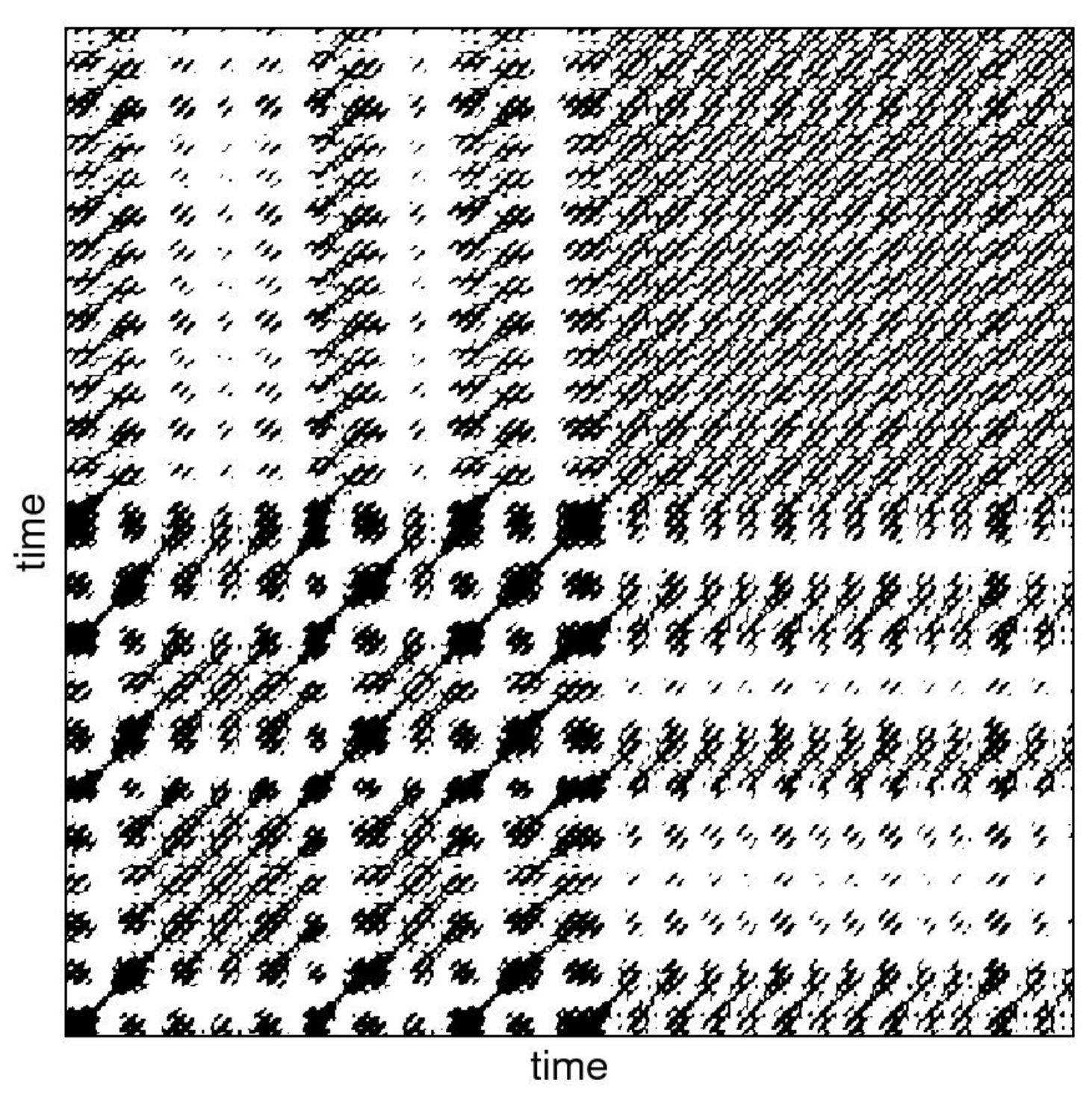}~~~
\includegraphics[scale=0.295,trim=0mm 0mm 0mm 0mm,clip]{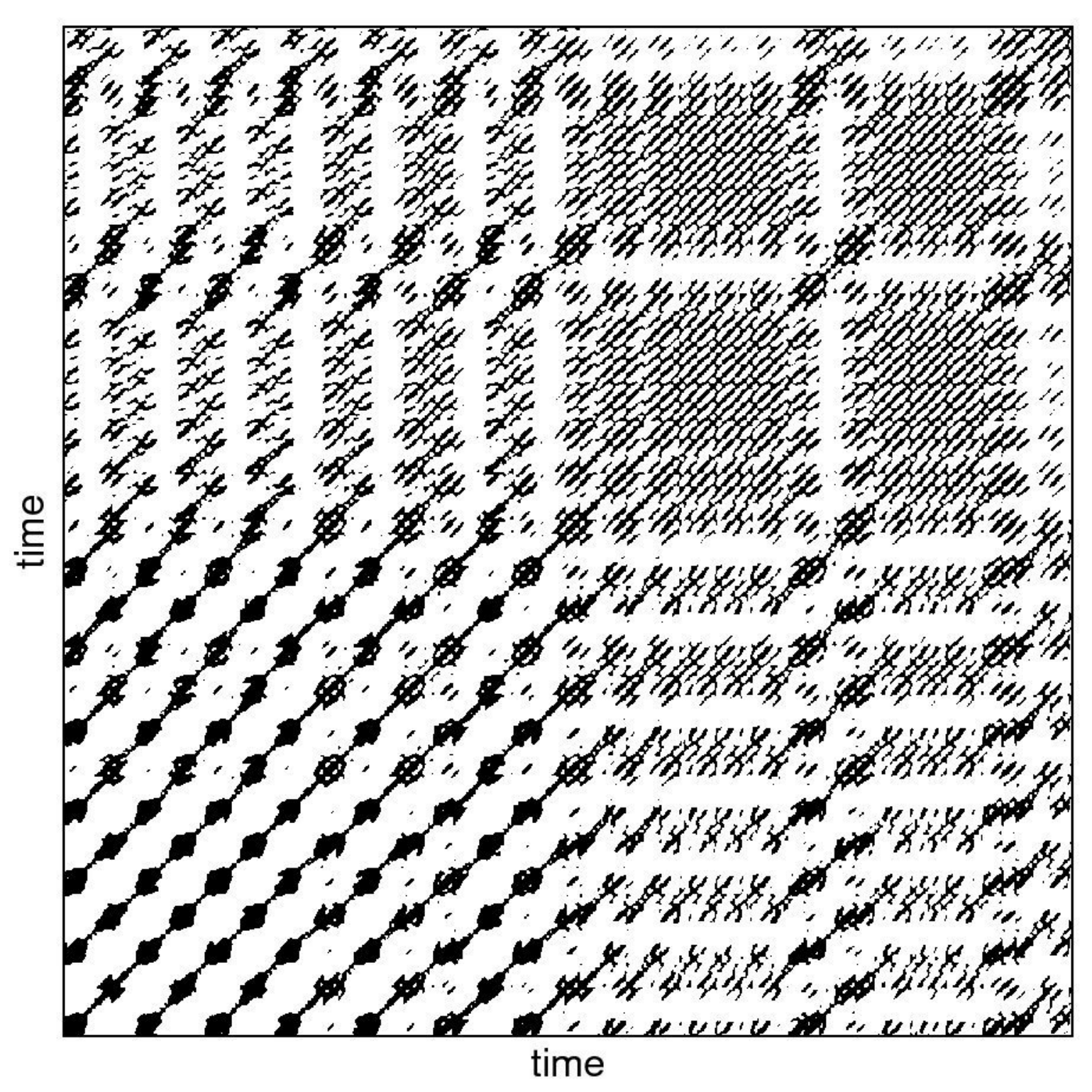}
\caption{Poincar\'{e} surfaces of section ($\theta_{\rm sec}=\pi/3$) and corresponding recurrence plots for charged particle with $q=5.581$ moving within halo lobes in the Bonnor spacetime with $b=1\,a$. Particle with $L=-2.356\,a$ is launched just from the locus of the off-equatorial potential well minimum ($r_{\rm min}=6\,a$, $\theta_{\rm min}=\theta_{\rm sec}=\pi/3$ and $V_{\rm min}=0.81675$) with various values of energy (from the upper left to the bottom right:  $E=0.8168$, $E=0.818$, $E=0.8182$, $E=0.8183$, $E=0.819$ and $E=0.8198$). First three couples of plots show the situation in the halo lobe, while bottom plots reveal the dynamics after merging of the lobes. A decisive surface of section cannot be constructed for the particle in a opened lobe ($E=0.8198$) as it escapes after several intersections with the surface, while the corresponding RP shows the chaotic nature of the motion unambigiously. Unlike the previous figures, we plot all intersection points: downward crossings with $u^{\theta}\geq0$ (black dot) as well as those resulting from upward crossings with $u^{\theta}<0$ (red dot) on the Poincar\'e surfaces.}
\label{Fig:13}
\end{figure} 

First panels of both figures show the situation if we set the energy $E$ to correspond to a small halo lobe, i.e., the energy of the particles is only slightly higher than that of corresponding circular halo orbit. In such a case, both the uncharged and charged particles show ordered motion - Poincar\'e sections are filled mainly with KAM tori and only one deformed resonant island appears. No traces of chaotic dynamics are present at all.  

When we increase the energy to the level corresponding to a large halo lobe which almost touches its symmetric twin below the equatorial plane, we obtain considerably more complex picture on the surfaces of section (upper right panels of figures \ref{Fig:11} and \ref{Fig:12}). In the uncharged case, we observe well pronounced Birkhoff chains of resonant islands to dominate the scene over receding KAM tori. On the other hand, considerable fraction of charged particles under analogous circumstances shows chaotic behaviour. Islands of stability are still present, but original KAM tori already disintegrated. 

Once the energy is further increased, the two symmetric halo lobes merge into one cross-equatorial region. The dynamics of uncharged particles changes dramatically, KAM tori mostly disappear and, only several islands of stability survive the onset of chaos (bottom left panel of figure \ref{Fig:11}). However, such abrupt change of the size and shape of potential lobe does not considerably affect the motion of charged particles for which chaos already appeared inside the halo lobe. The bottom left panel of figure \ref{Fig:12} reveals that the only difference is that few more islands of stability decompose. Increasing the cross-equatorial lobe to its maximal spatial extent causes decomposition of most islands of stability, though some of them resist. Surprisingly, more stable islands are then found for charged particles (bottom right panel of figure \ref{Fig:12}), while for the uncharged particles we observe only few traces of stable motion in the chaotic sea (bottom right panel of figure \ref{Fig:11}). 

We conclude that the energy of particle acts as a trigger for chaotic motion, which gradually shifts the dynamics from regular to chaotic. In particular, we confirm that motion of uncharged particles in Bonnor spacetime is nonintegrable, which was not clear enough from the discussion of equatorial lobes (figure \ref{Fig:10}). More strongly perturbed motion of ionised particles appears to be more prone to chaos, as it shows well inside the off-equatorial lobe, while uncharged particles do not become largely chaotic before merging both symmetric lobes, which allows them to cross the equatorial plane freely.

To illustrate the continuous transition from ordered to chaotic dynamics, we pick a particular trajectory of charged particle and plot series of Poincar\'e sections along with corresponding recurrence plots in figure \ref{Fig:13}. We launch the particle from the locus of the off-equatorial potential minimum with various values of energy $E$, while other parameters remain fixed. The sequence begins with the energy corresponding to a small halo lobe where we observe ordered motion manifested by narrow curves on the surface of section and, simple diagonal pattern of the recurrence plot (upper left panels of figure \ref{Fig:13}).  Increasing the energy, however, gradually shifts the dynamics towards deterministic chaos -- trajectory becomes more and more ergodic as it spans larger fraction of given energy hypersurface in the phase space. 

Recurrence plots help to reveal abrupt change of dynamics which occurs when the both symmetric halo lobes merge in a single cross-equatorial region. Indeed, we observe that RP attains typical chaotic pattern of disrupted diagonal lines when the lobes merge (compare upper right couple of panels to bottom left couple in figure \ref{Fig:13}). Bottom middle panels corresponding to large cross-equatorial lobe confirm the full onset of chaotic dynamics as the trajectory ergodically fills all allowed region on the surface of section, and diagonal segments of RP further disintegrate. The bottom right panel corresponds to the opened cross-equatorial lobe from which the particle escapes after certain amount of time which is too short for a construction of a decisive Poincar\'e surface of section, though, it is long enough to create an unambiguous recurrence plot of chaotic nature. We note that the integration time, which was necessary for the construction of presented surfaces of section, was typically around $100$ times longer than that we needed for plotting corresponding RPs.

The role which the quadrupolar deformation (\ref{Quadrupole}) plays in inducing the chaotic behavior has already been studied in various context in number of works. Guer\'{o}n and Letelier first considered the problem in the framework of classical Newtonian gravity \cite{gueron01} and, then also for exact solutions to Einstein equations \cite{gueron02}. In the latter paper the authors study both the static case of `distorted' Schwarzschild black hole as well as a rotating black hole deformed by a quadrupolar term. The solution they employ in the study of a static source is a special case of solution \cite{quevedo89} that incorporates the whole series of higher multipoles.  The rotating object is described by a metric obtained as a two-soliton solution using the Belinsky-Zakharov method \cite{belinsky79}. From the numerical study of test particle trajectories in all these systems they conclude that only the prolate quadrupole deformation ($\mathcal{Q}>0$) induces chaotic dynamics. Since they find no traces of chaos for oblate deformations ($\mathcal{Q}<0$), they suggest that the system could become integrable in this case \cite{gueron02}.

The issue has been revisited by Dubeibe et al. \cite{dubeibe07} who employ the solutions which are in a close relationship to Bonnor's massive dipole which we analyze in this paper. Namely, they study a test particle dynamics in Tomimatsu-Sato $\delta=2$ spacetime \cite{tomimatsu72} and its generalization found by Manko et al. \cite{manko00}. In the static limit, the former one reduces to Zipoy-Voorhees $\delta=2$ spacetime which is identical to the non-magnetized Bonnor spacetime obtained by setting $b=0$ in (\ref{metric}). On the other hand, the solution by Manko et al. generalizes the Bonnor's one by considering rotation and quadrupole moment as extra free parameters, while Bonnor's solution is static and  quadrupole moment is fixed by parameters $a$ and $b$ (\ref{Quadrupole}). The analysis by Dubeibe et al. leads to the conclusion that there are no chaotic orbits in Tomimatsu-Sato $\delta=2$ spacetime (whose deformation is always oblate) and the system might be fully integrable. On the other hand, the Manko et al. solution admits chaos only for oblate deformations. The last point has been contradicted by Han \cite{han08} who argues that the chaotic regions found by Dubeibe et al. for oblate deformations are not astrophysically relevant, since they only appear as narrow zones so close to the center that they would actually collide with the star. Han's analysis rather shows that motion around oblate Manko et al. star is typically regular, while the prolate deformation induces chaotic dynamics in large regions outside the star.

Both Dubeibe et al. and Han considered the Manko et al. solution without the magnetic dipole, being interested in a real deformation of the mass rather than additional effective quadrupolar deformation caused by the electromagnetic field. Han only briefly comments that including the magnetic dipole does not change the global dynamics and, in particular, that it does not bring chaos in oblate case. Only neutral test particles have probably been considered.

Our results cannot be directly contrasted with the works mentioned above, since we mainly study the dynamics in a slightly different setup (charged test particles and nonzero magnetic dipole). Nevertheless, we observe that in the case of neutral particles and oblate effective quadrupole moment (\ref{Quadrupole}), i.e., for $0\leq{}|b|<a$, the dynamics is typically regular as we found no chaos in closed lobes under these circumstances. The chaos was present only in very narrow zones formed by escaping orbits in opened lobes (see figure \ref{Fig:10b}). However, these zones were found well above the horizon of the central body. From this we conclude that although the chaos was extremely rare, the system is not integrable even for $b=0$ (see also recent analysis by Lukes-Gerakopoulos \cite{lukes12}). On the other hand, if we consider charged particles around Bonnor's source with oblate deformation $0<|b|<a$, we find large regions of chaos easily. Higher values of dipole parameter $|b|>a$ leading to prolate deformation ($\mathcal{Q}>0$) admit chaotic dynamics for both neutral and charged test particles. However, the role of effective quadrupole moment $\mathcal{Q}$ as a trigger for chaos is not clear from our results, since charged particles exhibit the chaotic motion even for $\mathcal{Q}=0$ ($|b|=a$).

\section{Conclusions}
Our aim with this paper was to study and compare different approaches towards the properties of motion near a magnetized compact body, taking 
into account the interplay of gravitational and electromagnetic effects that act on electrically charged matter. To this end we adopted the 
example of Bonnor's solution for the magnetic dipole as a spacetime metric that is given in an analytical (suitably simple) form, while 
preserving the essential features of a massive dipole. Our results are described in several sections which help us to proceed through different 
aspects in a systematic manner.

Firstly (section 3), we investigated properties of equatorial circular motion and, determined the regions of radius accessible for different 
combinations of the spacetime parameters and the particle constants of motion. Then we explored circular motion outside the equatorial plane (section 4). This allowed us to reveal the formation of the lobes of stable circulation, which typically occurs when the dipole-type magnetic field and the 
central gravitational attraction act simultaneously on charged particles. In comparison with the previously studied cases of halo circular orbits circling around compact objects with the presence of the dipole-type magnetic fields, the Bonnor's massive dipole permits the existence of even circular halo geodesics.  

In the following parts of the paper (sections 5--7), we presented mainly a numerical study of the transition between regular and chaotic motion. We analysed the motion in the equatorial potential wells (section 6) in three different cases, namely motion in non-magnetized spacetime with $b=0$, motion of uncharged particle on the magnetized background ($q=0$, $b\neq 0$) and dynamics in the general case $q\neq 0$, $b\neq 0$. Then, we again proceeded to the rich variety of properties of motion in the off-equatorial potential wells (section 7). 

Our results show that without magnetic field the system hosts mostly regular orbits and the dynamics of test particles resembles fully integrable systems. However, further numerical inspection revealed that chaotic orbits are also present in this setup proving the system nonintegrable. Then we observed that the magnetic parameter $b$ introduces a profound perturbation of the dynamics. Moreover, charge of the particle acts as an extra perturbation, which shifts magnetized system even farther from the integrability. We also studied the role of particle energy $E$ on the degree of chaos found in the system, concluding that it acts as a trigger for the chaotic motion. As the energy is gradually increased, the system undergoes a continuous transition from the regular behaviour to the chaotic dynamics, being almost fully ergodic on the given hypersurface. We illustrated such transition by means of Poincar\'e surfaces of section and recurrence plots. The role of quadrupole moment has also been discussed.

Because of the nature of Bonnor spacetimes, the contribution of presented results is more or less in theoretical level. However, the motion of charged particles within the off-equatorial lobes itself represents also very interesting astrophysical problem \cite{Kovar10} being, e.g., an essential step in modelling of circling off-equatorial structures around relevant compact objects. Such a work is being finished these days as well, describing electrically charged perfect fluid tori \cite{Kovar11} `levitating' above the equatorial plane in spherically symmetric and dipolar magnetic fields.

\ack
JK thanks for the support from the project of Czech Science Foundation GA \v{C}R ref.\ P209/10/P190 and from the project Synergy CZ.1.07/2.3.00/20.0071 supporting the international collaboration of Institute of Physics, Silesian University in Opava; the author would also like to express his acknowledgement for the Institutional support of Faculty of Philosophy and Science, Silesian University in Opava.
OK is grateful to obtain support from GA \v{C}R ref.\ 202/09/0772. VK acknowledges the continued support for the Czech-US collaboration project ME09036. The present work was started during the stay of YK in Prague, which was supported by Joint Research Program between Academy of Sciences of Czech Republic and Japan Society for the Promotion of Science. Astronomical Institute is supported by Academy of Sciences of Czech Republic (RVO 6798815).

\end{document}